\documentclass[reprint,secnumarabic,amssymb, nobibnotes, aps, prd, 
floatfix, showpacs]{revtex4-1}
\pdfoutput=1
\usepackage{graphicx} 
\usepackage{amsopn}
\usepackage{amsmath}
\usepackage{amssymb}
\usepackage{amsthm}
\usepackage{lmodern}
\usepackage{url}
\usepackage[colorlinks = true,
linkcolor = blue,
urlcolor  = blue,
citecolor = blue,
anchorcolor = blue]{hyperref}
\usepackage{bbm} % for representation of Unit-Matrix, use command: \mathbbm{1}
\usepackage[version=3]{mhchem}
\usepackage{rotating}
\usepackage{dsfont} %  unitmatrix, zeromatrix
\usepackage[usenames, dvipsnames]{color}
\usepackage[inline]{enumitem} % to have items enumerated in a line

\theoremstyle{plain} % default
\newtheorem*{theorem*}{Theorem}
\newtheorem*{remark}{Remark}

\newcommand{\executeiffilenewer}[3]{%
	\ifnum\pdfstrcmp{\pdffilemoddate{#1}}%
	{\pdffilemoddate{#2}}>0%
	{\immediate\write18{#3}}\fi%
}

\begin{document}
\title{Analytical~methods for~vacuum~simulations in~high~energy~accelerators for~future~machines based~on~the~LHC~performance}
\author{Ida Aichinger}
\email{ida.aichinger@cern.ch}
\affiliation{CERN, The European Organization for Nuclear Research, CH - 1211 Geneva, Switzerland}
\author{Roberto Kersevan}
%\email{roberto.kersevan@cern.ch}
\affiliation{CERN, The European Organization for Nuclear Research, CH - 1211 Geneva, Switzerland}
\author{Paolo Chiggiato}
%\email{paolo.chiggiato@cern.ch}
\affiliation{CERN, The European Organization for Nuclear Research, CH - 1211 Geneva, Switzerland}

\pacs{02.70.-c,	47.45.-n, 02.30.Hq, 29.20.db}
\date{June, 2017}

\begin{abstract}
The Future Circular Collider (FCC), currently in the design phase, will address many outstanding questions in particle physics. The technology to succeed in this 100~km circumference collider goes beyond present limits. Ultra-high vacuum conditions in the beam pipe is one essential requirement to provide a smooth operation. Different physics phenomena as photon-, ion- and electron- induced desorption and thermal outgassing of the chamber walls challenge this requirement. 
This paper presents an analytical model and a computer code PyVASCO that supports the design of a stable vacuum system by providing an overview of all the gas dynamics happening inside the beam pipes.  A mass balance equation system describes the density distribution of the four dominating  gas species $\ce{H2}, \ce{CH4}, \ce{CO}$ and $\ce{CO2}$. An appropriate solving algorithm is discussed in detail and a validation of the model including a comparison of the output to the readings of LHC gauges is presented. This enables the evaluation of different designs for the FCC. 
\end{abstract}

\maketitle

\section{Introduction}
The computation of residual gas particle (RGP) density profiles in particle accelerators is an essential task to optimize beam pipes and vacuum system design. In the last two decades, some software have been developed \cite{ady2014introduction,rossi2004vasco}. They have been used for most of the high-energy accelerators presently in activity, including the Large Hadron Collider (LHC) at CERN.  

There exists several approaches to evaluate gas density profiles \cite{bird1976molecular, lafferty1987vacuum, brush2003kinetic}. The most general one would be to tackle directly the nonlinear integro-differential Boltzmann equation\cite{harris2004introduction}. However, the solution of the Boltzmann equation requires an important computational effort due to the complicated structure of the collision integral. To reduce the complexity of the computation, gas density profile calculations have been performed by probabilistic Monte Carlo simulations, either by Direct~Simulation Monte Carlo method~(DSMC) \cite{bird1976molecular} or, in a simpler way, by the test~particle simulation Monte Carlo method~(TPMC) \cite{nakhosteen2016handbook}. Among the latter methods, MolFlow+ \cite{ady2014introduction} is largely spread in the vacuum technology community. However, although such methods have found applications beyond particle accelerators \cite{ady2016monte}, their extensions to time-dependent behaviours and multiple gas species phenomena are difficult, in particular for long vacuum sectors.

Analytical models can overcome such obstacles if the evaluation of one-dimension gas-density profiles is sufficient, preferably in simple geometries with cylindrical symmetry. A typical example is VASCO \cite{rossi2004vasco, bregliozzi2012vacuum}, which was developed at CERN in 2004 for the interaction regions of the LHC.
Recently, the preliminary design-study of the Future Circular Colliders~(FCC) \cite{benedikt2014future} with unprecedented high energy and vacuum requirements offers to extend the application of analytical methods.  As an example, the hadron-hadron version of the FCC with around 100~km circumference and 100~TeV centre of mass energy, requires a gas density in the arcs that is five times lower than the one in the LHC. In this paper, we revise and update the previous models, present the underlying theory and introduce a new elaborated software PyVASCO.
In this new analytical method, we combined multiple effects due to material outgassing, beam induced desorption, conductance limitations, and different pumping mechanisms. A coupled differential equation system describes the mathematical framework of the model. Each equation represents the mass balance of one of the four dominant gas species $\ce{H2}, \ce{CH4}, \ce{CO}$ and $\ce{CO2}$. These equations are coupled due to interaction of the different gas species among each other. For example, $\ce{CO2+}$\textendash after ionization by the beam\textendash may desorb $\ce{H2}$ from the beam pipe materials.
Mathematically, the problem translates into a large sparse matrix equation system of first order, as~in~\cite{rossi2004vasco}. We developed a new optimized solving algorithm and implemented the model in a Python environment. This resulted in a significantly improved performance in speed and memory storage allowing to simulate 100~times~longer vacuum systems than those achievable by the previous work \cite{rossi2004vasco} in less than 30~seconds.
We benchmarked the simulation output with MolFlow+ and cross checked it to the readings of pressure gauges installed in accelerators. The latter verification is presented for the Long~Straight~Section~(LSS)~4~and~5 of the LHC with a total length of over 1000~m along the beam line. \\
The focus of the simulation in this paper lies on circular hadron accelerators, since LHC's gauge reading are available for verification and additionally the FCC-hh presents the ultimate goal of the design study. Nevertheless, the code PyVASCO is applicable to any other type of particle accelerators, like lepton machines, linacs and heavy-ion accelerators.

\section{Setting up the physical vacuum model} \label{physical_model}

\begin{figure}[b]
	\centering
	\includegraphics[width=0.4\textwidth]{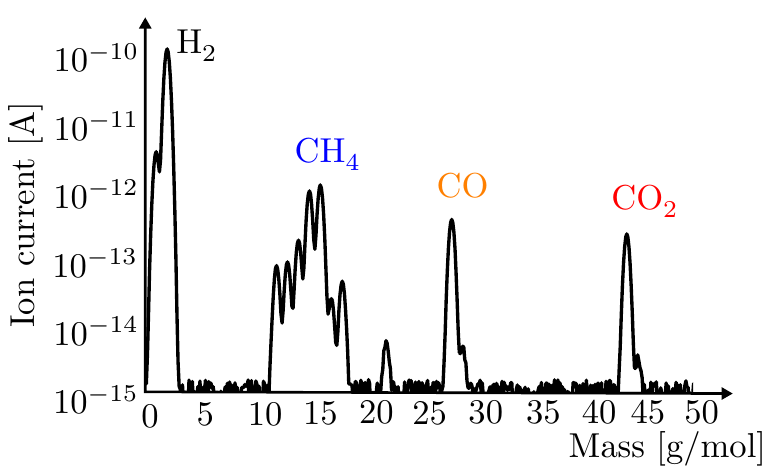}
	\caption{Mass spectrum in a UHV system. The four peaks correspond to the presence of  $\ce{H2}, \ce{CH4}, \ce{CO}$ and $\ce{CO2}$ gas particles. The measured ion-current can be converted to the gas density with appropriate calibration coefficients.}
	\label{fig:massSpectrum}
\end{figure}

This section provides an introduction to the physical quantities and laws that  form the equation system of the vacuum dynamics with the particle density $\mathbf{n}$ as unknown. The variables introduced here are summarized at the end of this section in TABLE \ref{PhyDescrTable}.\\
Mathematically, $\mathbf{n}$ represents the vector-valued density function of the four dominating gas species $\ce{H2}, \ce{CH4}, \ce{CO}$ and $\ce{CO2}$  as observed in a mass spectrum for a ultra-high vacuum (UHV) system (see Fig.~\ref{fig:massSpectrum}).
\begin{eqnarray}
\mathbf{n} = (n_{\ce{H2}}, n_{\ce{CH4}}, n_{\ce{CO}}, n_{\ce{CO2}})^T
\end{eqnarray}

The framework of the Frenet-Serret  coordinate system~\cite{kuhnel2015differential} is an appropriate choice for circular colliders, and the Cartesian coordinate system for straight linear colliders. In any case, we will refer to the vector pointing in the direction of the beam as $x$. The horizontal and vertical coordinates are noted by $y$ and $z$, the time is noted by $t$.
UHV systems are characterized by a high Knudsen number $Kn > 10$ with: 
\begin{eqnarray}
Kn = \frac{k_B T}{\sqrt{2}\pi \delta^2 p d},
\end{eqnarray}
where $k_B$ is the Boltzmann constant, $T$ the temperature, $d$ the beam pipe diameter, $\delta$ is the particle hard shell diameter and $p$ the total pressure. \\ 
The pressure $\mathbf{p}$ of the gas and the particle density $\mathbf{n}$ in a UHV-system are correlated via the ideal gas law: 
\begin{eqnarray}
\mathbf{p} = \mathbf{n} \cdot k_B \cdot T
\end{eqnarray}
For clarification, the total pressure is the sum of the partial pressure values of each gas specie $i$:
$$ p = \sum\limits_{i=1}^4 \mathbf{p}_i $$
The Maxwell Boltzmann distribution describes the particle speed $v$ for ideal gases \cite{nakhosteen2016handbook, chapman1970mathematical}. %p. 38
The corresponding mean velocity is given by
\begin{eqnarray}
\overline{\mathbf{v}} = \sqrt{\frac{8 k_B T}{\mathbf{m} \pi}} \label{velo}
\end{eqnarray}
depending on the molecular mass $\mathbf{m}$ and $T$.
Fick's first and second laws of diffusion \cite{crank1979mathematics, bird1976molecular} define the fundamental balance equation of the gas kinetics in a UHV-system. We get the following equation for an isotropic medium with a constant diffusion coefficient $\mathbf{a}$:
\begin{eqnarray} \label{balance}
\frac{\partial \mathbf{n}}{\partial t} = \mathbf{a} \frac{\partial^2 \mathbf{n}}{\partial x^2}
\end{eqnarray}
if diffusion is one-dimensional.  
This assumption is appropriate since most of the time the length of the vacuum chamber is much bigger than its cross section, thus diffusion occurs mainly along the beam line. Furthermore, experimental data from the laboratory is also one-dimensional since they measure the gradient of gas concentration along the x-axis.

The diffusion coefficient $\mathbf{a}$ depends on the particle's speed, its mass, and on the beam pipe geometry. Under molecular flow conditions, $\mathbf{a}$ is given for cylindrical vacuum chambers by the specific conductance based on the calculations of Knudsen \cite{knudsen1909gesetze, nakhosteen2016handbook}:
\begin{eqnarray}
\mathbf{a} = \Big(\frac{d}{2}\Big)^2 \pi \cdot  \frac{ d}{3} \cdot \overline{\mathbf{v}} = \frac{d^3 \pi }{12} \cdot \overline{\mathbf{v}}
\end{eqnarray}
In order to generate a correct mass balance, continuous flow into and out of the system (see also subsection \ref{flow_into}, \ref{flow_out}) , e.g. in terms of desorption or pumping, at a rate $\mathbf{q} $ and $\mathbf{r} $ per unit volume, must be added to the right side of Eq. \eqref{balance}:
\begin{eqnarray}
\frac{\partial \mathbf{n}}{\partial t} = \mathbf{a} \frac{\partial^2 \mathbf{n}}{\partial x^2} + \mathbf{q}(x,t) - \mathbf{r}(x,t) \cdot \mathbf{n} \label{equSys}
\end{eqnarray}
Note, that the flow out of the system is always proportional to the prevailing density $\mathbf{n}$, whereas the flow into the system can act independently of $\mathbf{n}$.
Local sinks and sources, e.g. due to lumped pumps or outgassing related to possible tiny leaks are considered in the  boundary conditions. 

\begin{figure}[b]
	\centering
	\includegraphics[width=0.4\textwidth]{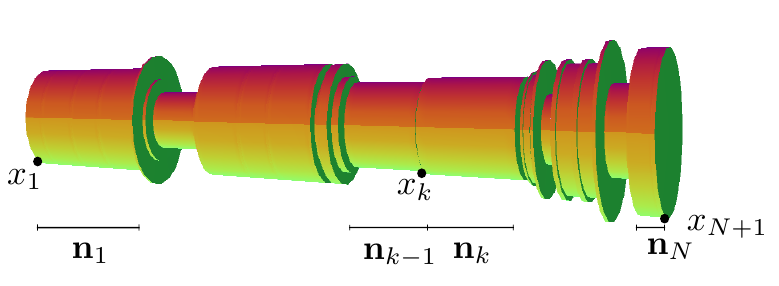}
	\caption{Segmenting domain in $N$ parts. Solution $n_k$ describes density on segment $k$.}
	\label{fig:segmenting}
\end{figure}

The idea is now to fragment the domain into a finite number of elements, where $ \mathbf{q} $, $ \mathbf{r} $ and $ \mathbf{a} $ can be taken as constant vectors. This allows to solve the equation system \eqref{equSys} locally on each element. The piecewise solution concept is identified by an additional index $k$ (see~Fig.~\ref{fig:segmenting}).

Accordingly, appropriate intersection conditions combine the solutions to a global model. 
The conservation principle applied at the interface between two elements $k$ and $k+1$  leads to density and flux continuity conditions. Thus, it holds at the intersection point $x_k$ with a lumped gas source $\mathbf{g}$ and a lumped pump $\mathbf{s}$ (ion-pumps, turbomolecular pumps) that:
\begin{eqnarray}
 \mathbf{n}_{k-1}   &=&  \mathbf{n}_k \label{boundary_cond_IC1} \\
- \frac{D\mathbf{n}_{k-1}}{Dt} + \frac{D\mathbf{n}_{k}}{Dt}  &=&  \mathbf{s}_k \mathbf{n}_k  - \mathbf{g}_k \label{boundary_cond_IC2} , 
\end{eqnarray}
where the total derivative for a space-and-time dependent variable is given by \cite{bird2007transport, ferziger1973mathematical}:
\begin{eqnarray}
\frac{D\mathbf{n}}{Dt} := \frac{\partial \mathbf{n}}{\partial t} + \mathbf{a} \frac{\partial \mathbf{n}}{\partial x}  
\end{eqnarray}
Note, the change of sign in front of the term involving the total derivative: The derivative at the beginning of segment $k$ points in the positive direction, while at the end of segment $ k-1 $, it points in the negative direction.\\

At the extremities of the vacuum chamber open boundary conditions are assumed, hence only half of the pumping speed and half of the gas release is considered. 
\begin{eqnarray}	
\frac{D \mathbf{n}_1}{Dt} = \frac{\mathbf{s}_1}{2} \mathbf{n}_1  - \frac{\mathbf{g}_1}{2}  \qquad &\textrm{for }& x = x_0\\
\frac{D \mathbf{n}_N}{Dt} = \frac{\mathbf{s}_{N+1}}{2} \mathbf{n}_N  - \frac{\mathbf{g}_{N+1}}{2}   \qquad  &\textrm{for }& x = x_{N+1}
\end{eqnarray}
Note, also periodic boundary conditions at the extremities present a possible choice that set $\mathbf{n}_1 $ and $\mathbf{n}_N $ in the sense of \eqref{boundary_cond_IC1} and \eqref{boundary_cond_IC2} in relation.  

\subsection{Flow into the vacuum-system $\mathbf{q}$} \label{flow_into}

There are four main phenomena, where particles are added to the system. 
The first three phenomena, described in the following subsections, are due to beam dynamic effects and are considered as the main impact on the dynamic gas density. In these cases, energetic particles as photons, ions and electrons bombard the chamber. If the energy spectrum of the impinging particles is in the  electronic structure of the beam pipe material, then molecular desorption from the chamber walls may take place. The desorption probability is described by the parameter $\eta$. Its value is mainly specified by experimental data. 
In general, we consider three different types of impinging particles as photons, ions and electrons which lead to different desorption phenomena:

\subsubsection{Photon-induced desorption}  \label{flowinto1}
Accelerated charged particles emit photons in the presence of magnetic fields. This process is called synchrotron radiation (SR) \cite{hofmann2004physics}. Its impact on vacuum systems is common for electron-positron circular colliders as for LEP \cite{bailey1998synchrotron} and could also be observed for high energy-proton ciruclar colliders as the LHC \cite{tuckmantel2005synchrotron}. The total number of photons emitted~per~second and per meter is described with the photon flux $\dot{\Gamma}$.  It depends on the energy of the accelerated particle expressed by the relativistic factor $\gamma$ and the strength of the magnetic field in terms of the bending radius $\rho$: 
\begin{eqnarray} \label{photonflux}
\dot{\Gamma} = \frac{5 \sqrt{3} \gamma}{24 \pi \epsilon_0 \hbar \rho c},
\end{eqnarray} 
where $\hbar$ is the reduced Planck constant, $c$ the speed of light and $\epsilon_0$ the  permittivity of vacuum.
The process of the photon-induced desorption depends on the energy spectrum of the impinging photons and on the material of the vacuum chamber. Nearly a linear dependence  has been observed between the critical photon energy $E_c$ in the range from 10-1000eV and the desorption yield for $\eta_{ph}$ for the most commonly used materials of the beam chambers \cite{gomez1994comparison}. 
$E_c$  is the median of the photon energy spectrum:
\begin{eqnarray*}
E_c = \frac{3}{2}\frac{\hbar c}{\rho} \gamma^3,
\end{eqnarray*}
The derived desorption yield for a copper-lined chamber, e.g. is given by \cite{gomez1994comparison}:
\begin{eqnarray*}
\eta(\ce{H_2}) \sim E_c^{ 0.74}, \quad \eta(\ce{CH_4}) \sim E_c^{ 0.94}, \\ 
\eta(\ce{CO}) \sim E_c^{ 1.01}, \quad \eta(\ce{CO_2}) \sim E_c^{ 1.12}
\end{eqnarray*}
Note, that photons below 4 eV do not provoke desorption, consequently the previous formula is valid, when the integral up to 4 eV presents only a small fraction of the total integral.  In \cite{gomez1994comparison}, the photon induced desorption was studied for a $E_c$ up to 300eV.

\subsubsection{Ion-induced desorption} \label{flowinto2}

The beam can ionize the RGP \cite{turner1996ion}.  Consequently the generated positive ions are repelled by the positive space charge of the proton beam and imping on the vacuum chamber walls, where they may cause desorption of tightly bounded molecules from the surface. This phenomena was observed for example in the Intersecting Storage Ring (ISR) at CERN \cite{calder1974ion}. 

In this process the different gas species may influence each other.
The interaction is described by the desorption yield matrix 
\footnotesize
\begin{eqnarray*}
	\mathbf{H}_{\textrm{ion}} = \left(
	\begin{array}{llll}
		\eta_{\ce{H2+} \rightarrow \ce{H2}} & \eta_{\ce{CH4+} \rightarrow \ce{H_2}} &  \eta_{\ce{CO+} \rightarrow \ce{H2}} & \eta_{\ce{CO2+} \rightarrow \ce{H2}} \\
		\eta_{\ce{H2+} \rightarrow \ce{CH4}} & \eta_{\ce{CH4+} \rightarrow \ce{CH4}} &  \eta_{\ce{CO+} \rightarrow \ce{CH4}} & \eta_{\ce{CO2+} \rightarrow \ce{CH4}} \\
		\eta_{\ce{H2+} \rightarrow \ce{CO}} & \eta_{\ce{CH4+} \rightarrow \ce{CO}} &  \eta_{\ce{CO+} \rightarrow \ce{CO}} & \eta_{\ce{CO2+} \rightarrow \ce{CO}} \\
		\eta_{\ce{H2+} \rightarrow \ce{CO2}} & \eta_{\ce{CH4+} \rightarrow \ce{CO2}} &  \eta_{\ce{CO+} \rightarrow \ce{CO2}} & \eta_{\ce{CO2+} \rightarrow \ce{CO2}} \\
	\end{array}
	\right).
\end{eqnarray*}
\normalsize{}
For clarification, the entry $H_{4,1} = \eta_{\ce{CO2+} \rightarrow \ce{H2}}$ describes the probability that a $\ce{CO2+}$ ion desorbs a $\ce{H2}$ molecule. 

The process of desorbed particles is thus described with the product of the desorption yield matrix $ \mathbf{H}_{\textrm{ion}} $ and the ion flux $ \dot{\mathbf{I}}_{\textrm{ion}} $:
\begin{eqnarray*}
	\mathbf{H}_{\textrm{ion}} \cdot \dot{\mathbf{I}}_{\textrm{ion}} \cdot \mathbf{n} = \sum\limits_{j=1}^4\mathbf{H}_{i,j}  \cdot \frac{I}{e} \sigma_j \cdot \mathbf{n}_{j} ,
\end{eqnarray*}
where the summation runs over all impinging species $j$ that desorb particles of species $i$. The summand describes the product of the ion-induced desorption matrix entry $ (H_{i,j}) $ with the number of ionized particles $(\frac{I}{e} \sigma_j \cdot \mathbf{n}_{j})$, where $I$ is the beam current, $e$ the elementary charge, $\sigma$ the ionization cross section and $\mathbf{n}$ the RGP density. The ionization cross section for energies greater than 100 keV for a gas particle is calculated in \cite{mathewson1996beam}. 
Additionally, it is worth mentioning that $\sigma$ is also depending on the ion energy, which in turn, depends on the beam current, making the ion-induced desorption term quadratic dependent on the beam current.  

\subsubsection{Electron-induced desorption - Electron cloud} \label{flowinto3}
The beam can generate some electrons from  synchrotron radiation,  impinging ionized gas molecules and spontaneous desorption induced by sufficiently high electromagnetic fields \cite{iadarola2014electron}. These primary electrons are accelerated by the positively charged beam, that hit the chamber wall and may produce a cascade of secondary electrons. The electrons are accelerated towards the positively charged beam, they cross the chamber and hit the wall again at the opposite side - producing more electrons and lead eventually to beam instabilities and gas desorption. This phenomenon depends in a complex way on the beam and chamber parameters and also on the bunch filling pattern.  A CERN proprietary software PyECloud~\cite{iadarola2013pyecloud}  addresses this phenomena. 
Based on observations, we can conclude that the bigger the aperture of the vacuum chamber, the longer is the duration that an electron is accelerated, the bigger is the surface  where desorption can take place and the higher is the electron induced desorption. In addition, it has been observed in the LHC that the reduction of the bunch spacing to 25ns causes a significantly increased electron cloud effect in comparison to 50ns or to 100ns~\cite{bradu2016compensation}.

\subsubsection{Other sources and summary of the total flow into the system} \label{flowinto4} 
Thermal desorption generates what is usually called the static vacuum, which is present even in the absence of the beam. For example, chamber walls and the components within the chamber randomly release gas which was adsorbed at the surfaces or entrapped into the bulk of the material. Air and water vapour may enter the system through leaks or permeate through seals. Gauges and beam instrumentation provide an additional source of outgassing.\\

Summarizing the flow into the vacuum system, gives the following expression for $\mathbf{q}$:
\begin{eqnarray*}
\mathbf{q}(x,t) &=&  \underbrace{	\mathbf{H}_{ion} \cdot \dot{I}_{ion} \cdot \mathbf{n} }_{\textrm{ion-induced desorption}}
+  \underbrace{\mathbf{\eta}_{ph} \cdot \dot{\Gamma}_{ph}}_{\textrm{photon-induced desorption}}  \\ 
&&
+  \underbrace{\mathbf{\eta}_{e} \cdot \dot{N}_{e}}_{\textrm{electron-induced desorption}} + 
\underbrace{\pi \cdot d  \cdot \mathbf{q}_{\textrm{th}}}_{\textrm{thermal outgassing}} 
\end{eqnarray*}
\subsection{Flow out of the vacuum-system $\mathbf{r}$} \label{flow_out}
Particles are continuously added to the UHV-system, as we have seen in the previous section. Therefore sufficient pumping systems have to be installed, in terms of distributed and spatially localized pumps. The latter one is mainly provided by conventional pumps e.g. ion pumps or turbomolecular pumps. 
Apart from this, impinging RGP may also stick on the wall due to thermodynamic or chemical binding at the surface of the chamber. This describes then distributed pumping. For example, a special surface coating of the vacuum chamber of $ \ce{Al}, \ce{Zr}, \ce{Ti}, \ce{V} $ and $ \ce{Fe} $, better known as Non-Evaporable Getter (NEG) \cite{benvenuti2001vacuum} induces distributed pumping. The RGP of $\ce{H2}$, $ \ce{CO} $ and $ \ce{CO2}$ are first chemically trapped by the NEG coating and then adsorbed into the bulk of the material. After a surface coverage of about one monolayer of adsorption of $ \ce{CO} $ and $ \ce{CO2} $ on the surface, the NEG saturates and the pumping efficiency drops down to negligible values. $\ce{CH_4}$ and noble gases are not adsorbed by NEG \cite{chiggiato2006ti}. \\
Distributed pumping also occurs in cryogenic areas, when a gas-particle hits the wall and immediately condenses. This phenomenon is known as cryo-pumping \cite{haefer2013kryo}.
Generally, the number of molecules impinging on the wall under molecular flow equilibrium conditions is given by 
\begin{eqnarray}
\frac{A \cdot \overline{\mathbf{v}} }{4},
\end{eqnarray}
where $ A $ describes the lateral surface of the vacuum chamber, and $\overline{\mathbf{v}}$ the mean velocity of the RGP  as defined in Eq.~\eqref{velo}. 
In contrast to the distributed pumping  particles sometimes get pumped only at an orifice on the beam pipe wall. For example, the holes in the LHC beam screen provide a linear pumping \cite{cruikshank1997mechanical}.  

\onecolumngrid
\vspace{\columnsep}

\subsection{Problem description} \label{problem_descr}
The central part of the model are the balance equations and the boundary conditions, summarized here: \\
Mass-Balance equation for segment $k$ (discarding here the index $k$ for each coefficient):
\begin{eqnarray} \label{balanceEQ}
\underbrace{ \frac{\partial \mathbf{n}}{\partial t}}_{\substack{\text{Time variation} \\ \text{of particles}}} =
\underbrace{\mathbf{a} \circ \frac{\partial^2 \mathbf{n}}{d x^2}}_{\substack{\text{Diffusion}}}+\underbrace{\mathbf{\eta}_{ph} \cdot \dot{\Gamma}_{ph}}_{\substack{\text{Desorption} \\ \text{by photons}}}
+ \underbrace{\mathbf{\eta}_{e} \cdot \dot{N}_{e}}_{\textrm{by electrons}} 
+ \underset{{\color{red}Multi gas model}} {\boxed{ \underbrace{ \mathbf{H}_{ion} \cdot  \dot{I}_{\textrm{ion}} \circ \mathbf{n}  }_{\substack{\text{Ionization by beam} \\ \text{and desorption by ions}}} }}
+ \underbrace{ A \cdot \mathbf{q}_{\textrm{th}}}_{\substack{\text{thermal} \\ \text{outgassing}}}
-  \underbrace{\alpha \circ  \frac{A \cdot \overline{\mathbf{v}} }{4} \circ (\mathbf{n}- \chi_{\textrm{cryo}} \mathbf{n_e}  )}_{\textrm{wall distributed pumping}} 
-  \underbrace{\mathbf{p}_{l} \circ \mathbf{n}}_{\substack{\text{linear} \\ \text{pumping}}} \qquad \quad
\end{eqnarray}
Boundary and intersection conditions for segment $k-1$ and $k$:
\begin{eqnarray} \label{IC}
 \mathbf{n}_{k-1}(x_k) &=& \mathbf{n}_k(x_k)  \\
-\mathbf{a}_{k-1} \circ \mathbf{n}'_{k-1} (x_k) + \mathbf{a}_{k} \circ \mathbf{n}'_{k} (x_k) &=& \mathbf{s}_k \circ \mathbf{n}_k (x_k) - \mathbf{g}_k  \\
\mathbf{a}_{1} \circ \mathbf{n}'_{1} (x_1) &=& \frac{\mathbf{s}_1}{2} \circ \mathbf{n}_1 (x_1) - \frac{\mathbf{g}_1}{2}\\
- \mathbf{a}_{N} \circ \mathbf{n}'_{N} (x_{N+1}) &=& \frac{\mathbf{s}_{N+1}}{2} \circ \mathbf{n}_N (x_{N+1}) - \frac{\mathbf{g}_{N+1}}{2}  
\end{eqnarray}
\begin{remark}
The symbol $ \circ $ indicates a component wise multiplication of two vectors. The symbol $'$ indicates the normal derivative at the boundaries.\\
Vectors are presented by lower-case bold letters (with the exception of Greek symbols) and matrices are presented by upper-case bold letters.\\
\end{remark}
\begingroup
%\squeezetable
\begin{table}[h]
	\caption{\label{PhyDescrTable} Model parameters for physical description.}
	\begin{ruledtabular}
		\begin{tabular}{llll}			
			\textbf{Symbol} & 
			\textbf{Dim} &
			\textbf{Unit}&
			\textbf{Description} \\
			\hline		
			$ \mathbf{n} $ & $\mathbb{R}^4$ &  Particles/$\textrm{m}^3$ & vector-valued particle density function of $ \ce{H2}, \ce{CH4}, \ce{CO} $ and $ \ce{CO2} $ \\
			$ \mathbf{a} $ & $\mathbb{R}^4$ & $\textrm{m}^4/s$ & Specific conductance \\
			$ \dot{\Gamma}_{ph} $ & $\mathbb{R}$&  $photons/ (s\cdot \textrm{m})$   & Emitted photon flux by the bended beam in the magnetic areas \\
			$ \dot{N}_{e} $ & $\mathbb{R}$&  $ electrons/(s\cdot \textrm{m})$   &  Electron flux impinging on the chamber wall due to the electron cloud phenomena \\
			$ \dot{I}_{\textrm{ion}} $& $\mathbb{R}$ &  $ ion/(s\cdot \textrm{m})$   &  Ion flux, it is proportional to n\\
			$ \frac{I}{e} $ & \quad  & $ \frac{1}{s} $ & Number of high energy protons passing per second \\
			$ \eta_{ph} $ &  $\mathbb{R}^{4}$  & 1 & Photon induced desorption yield $(\eta \geq 0 )$ describes the number of molecules desorbed per photon\\
			$ \eta_{e} $ &  $\mathbb{R}^{4}$  & 1 & Electron induced desorption yield $(\eta \geq 0) $ describes the number of molecules desorbed per electron\\
			$ \mathbf{H}_{ion} $ & $\mathbb{R}^{4 \times 4}$ &  1 & Ion induced desorption yield, probability that ion of specie $ i $ desorbs molecule of \\
			&&&specie $ j $ for $ i,j \in \{\ce{H2}, \ce{CH4}, \ce{CO}, \ce{CO2} \}$ \\
			A & $ \mathbb{R} $ &  m & Lateral surface per unit-length of beam chamber  \\
			$ \mathbf{r} $ & &   & Sinks of a UHV-system  \\
			$ \mathbf{q} $ & &  & Sources of a UHV-system  \\
			$ \mathbf{q}_{\textrm{th}} $ & $\mathbb{R}^{4}$ & $ 1/(\textrm{m}^2 s) \footnotemark[1] $  & Thermal outgassing rate \\
			$ \alpha $ & $\mathbb{R}^{4}$ & $ 1 $  & Sticking coefficient \\
			$ \overline{\mathbf{v}} $ & $\mathbb{R}^{4}$ & $ \textrm{m}/s $  & Average Maxwell-Boltzmann velocity of the four gas species  \\
			$ \mathbf{p}_l $ & $\mathbb{R}^{4}$ & $ \textrm{m}^2/s $  & Linear pumping per unit-length \\	
			N & $\mathbb{R}$ & 1 & Number of segments  \\
			d &  $\mathbb{R}$ & m & Diameter   \\
			L &  $\mathbb{R}$ & m & Length of segment  \\
			T &  $\mathbb{R}$ & K & Absolute temperature of segment  \\
			$ \mathbf{s}_k $ & $\mathbb{R}^{4}$ &  $ \textrm{m}^3/s $ & Pumping speed of lumped pump located on the beginning of segment $k$  \\
			x &  $\mathbb{R}$ &  m & Spatial coordinate along beam line\\
			$ x_k $ &  $\mathbb{R}$ &  m & Intersection point of segment $k-1$ and $k$\\
			$ \mathbf{g} $ &  $\mathbb{R}^{4}$ & $1/s$   & Local punctual gas source (e.g. gas leak) \\
			$ \sigma $ & \quad  & $\textrm{ion}/\textrm{proton} \cdot \textrm{m}^2 $ & Ionisation cross section of residual gas molecules by high energy protons \\	
			$ k_B  $&  \quad & $ \textrm{m}^2 kg/(s^2 K) $ & Boltzmann constant $ k_B $ = $ 1.3806488 \cdot 10^{-23} $\\
			$ p  $&  $\mathbb{R}$ & Pa & Total pressure\\
			$ \mathbf{p}  $&  $\mathbb{R}^{4}$ & Pa  & Equivalent pressure for particle density $\mathbf{n} $ using ideal gas equation\\
			$ \chi_{\textrm{cryo}}  $&  $\mathbb{N}$ & \quad  & $\chi = 1$ for cryogenic areas and 0 for room temperature areas\\
			$ \mathbf{n_e}  $&  $\mathbb{R}^4$ & $ 1/\textrm{m}^3 $  & Background density without beam (static density)\\ % typical value 1e10
		\end{tabular}
	\end{ruledtabular}
	\footnotetext[1]{[	$\widetilde{\mathbf{q}}] =[ \frac{\textrm{mbar} \cdot \textrm{l}}{\textrm{s} \cdot \textrm{cm}^2}] $ is more common in practice. $\mathbf{q}_{\textrm{th}} = \widetilde{\mathbf{q}} \cdot \frac{10^3}{k_B \cdot T}$ }
\end{table}
\endgroup
\twocolumngrid
	
\section{Analytical Solution Method}
The introduced physical description of a vacuum system provides us now a model, that needs to be solved. We focus here on the solution concept from a mathematical point of view. 
The differential Eq.~\eqref{balanceEQ}  has a solution that can be written in an exact and closed form under stationary assumptions. This assumption is applicable after a specific amount of pump-down time or when the accelerator operates at stable beam, then time variations are negligible and a stationary solution delivers accurate results.  
This leads to an elliptic partial differential equation with piecewise constant coefficients.\\
Introduced variables are summarized again at the end of this section in Table \ref{para_math}.\\
An equivalent stationary problem description of the balance equation is given by:
\begin{eqnarray}
\overrightarrow{0}_4 = \mathbf{A}(x) \frac{d \mathbf{n}^2}{dx^2} + \mathbf{B}(x) \mathbf{n} + \mathbf{c}(x) \label{ProblemD1}
\end{eqnarray}
with $  \mathbf{A}, \mathbf{B} \in \mathbb{R}^{4\times 4} $ and $ \mathbf{c} \in \mathbb{R}^4 $ being the matrix and vector assemblies of the parameters from the previous section:
\begin{eqnarray*}
\mathbf{A} &=& \mathbf{a} \cdot \mathbf{I}_{4}\\
\mathbf{B} &=& \mathbf{H}_{ion} \cdot  \dot{I}_{\textrm{ion}}- 
\mathbf{\alpha}  \circ \frac{\text{A} \cdot \overline{\mathbf{v}}}{4} - \mathbf{p}_l\\
\mathbf{c} &=& 
- \chi_{cryo} \cdot  \mathbf{\alpha}  \circ \frac{A \cdot \overline{\mathbf{v}}}{4} \circ n_e  +
\mathbf{\eta}_{ph} \cdot \dot{\Gamma}_{ph} + 
\mathbf{\eta}_{e} \cdot \dot{N}_{e} + 
A \cdot \mathbf{q}_{th} \label{Math_description}
\end{eqnarray*}
The major challenge in solving this system lies in $\mathbf{B}$. A fully occupied matrix $ \mathbf{B} $ couples the balance equation of each specie with each other. 
The idea is to transform Eq.~\eqref{ProblemD1} to a system of first order equations, for which a solution concept is known \cite{hsieh2012basic, crank1979mathematics, saff2015fundamentals}. Additionally, the idea is to split the domain into a finite number of segments as it was already done in the previous section (see also Fig.~\ref{fig:segmenting}), so that  Eq.~\eqref{ProblemD1} provides constant coefficients on each segment. We solve the equation system independently on each segment and connect the obtained solutions with transformed intersection conditions. Thus, we can finally formulate a global solution $\mathbf{n}$:
\begin{eqnarray}
\mathbf{n}(x) = 
\left\{ 
\begin{aligned}
\mathbf{n}_1(x)  & \quad  & x_1 &\leq x \leq x_{2} \\
\mathbf{n}_2(x)  & \quad & x_2 &< x \leq x_{3} \\
\vdots \quad   & \quad & \quad & \quad \\
\mathbf{n}_N(x)  & \quad &  x_N &< x \leq x_{N+1}
\end{aligned}
\right.
\label{globalSol}
\end{eqnarray}
\subsection{Transformed problem description}

The problem description \eqref{Math_description} is converted to a system of first-order linear equations with a change of variable. This modification reduces the order by one, but also doubles the amount of equations posed.
\begin{eqnarray*}
\mathbf{y} &:=& \begin{pmatrix} \mathbf{n} \\ \frac{d\mathbf{n}}{dx} \end{pmatrix} \\
\mathbf{M}   &:=& 
\left( \begin{array}{cc}
	\mathbf{0}_{4 \times 4} &  \mathbf{I}_{4} \\
	-\mathbf{A}^{-1} \mathbf{B} &  \mathbf{0}_{4 \times 4}
\end{array}\right) ,
\mathbf{b} := \begin{pmatrix}
	\mathbf{0}_{4 \times 4} \\
	-\mathbf{A}^{-1} \mathbf{c}
\end{pmatrix} \\
\mathbf{F_1}  &:=& \begin{pmatrix}
	-\frac{\mathbf{s}_{1} \mathbf{I}_{4}}{2} \quad \mathbf{A}
\end{pmatrix} ,
\mathbf{F_N} := \begin{pmatrix}
	-\frac{\mathbf{s}_{N+1}   \mathbf{I}_{4} }{2} \quad -\mathbf{A}
\end{pmatrix}
\end{eqnarray*}
\begin{eqnarray*}
\mathbf{H}_k &:=& 
\begin{pmatrix}
	\mathbf{I}_{4 } & \mathbf{0}_{4 \times 4} \\
	\mathbf{0}_{4 \times 4} & -\mathbf{A}_{k}
\end{pmatrix} , 
\mathbf{S}_k := 
\begin{pmatrix}
	\mathbf{0}_{4 \times 4} & \mathbf{0}_{4 \times 4} \\
	\mathbf{s}_k \mathbf{I}_{4 } & \mathbf{0}_{4 \times 4}
\end{pmatrix} \textrm{ and } \\
 \overline{\mathbf{g}}_k &:=& 
\begin{pmatrix}
	\overrightarrow{0}_4 \\
	-\mathbf{g}_k
\end{pmatrix}
\end{eqnarray*}

 The problem description reads now as follows: \\
\begin{eqnarray}
\frac{d\mathbf{y}}{d x}(x) = \mathbf{M} \mathbf{y}(x) + \mathbf{b} \label{balance1}
\end{eqnarray}
\begin{eqnarray}
\mathbf{H}_{k-1} \mathbf{y}_{k-1} (L) - (\mathbf{H}_k+\mathbf{S}_k) \mathbf{y}_k (0) &=& \overline{\mathbf{g}}_k \label{IC1}\\
\mathbf{F}_1 \mathbf{y}_1 (0) &=& -\mathbf{g}_1 \label{BC1}\\
\mathbf{F}_N \mathbf{y}_N(x_{N+1}) &=& \mathbf{g}_{N+1}, \label{EC1}
\end{eqnarray}
where $ \eqref{balance1} $  describes the balance equation, $ \eqref{IC1} $ the  intermediate condition, $ \eqref{BC1} $ the initial condition and $ \eqref{EC1} $ the end condition. 
\subsection{Solution for segment $k$}

The existence and uniqueness of a solution $ y(x) $ for segment $k$ to the Eq.~\eqref{balance1} with an arbitrary constant $u$ is posed by the fundamental Theorem of Picard Lindel\"{o}f \cite{picard1890memoire, lindelof1894application, lindelof1900demonstration}. The index $ k $ is again discarded for readability.
For each segment $k$, the solution $ y(x) $ can be stated as
\begin{eqnarray}
{\boxed{
		\mathbf{y}(x) = \underbrace{e^{(x-x_k) \mathbf{M}}}_{\mathbf{P}(x)} \mathbf{u} + \underbrace{\int_{x_k}^{x}e^{(x-\tilde{x}) \mathbf{M}} \mathbf{b} \;  d \tilde{x}}_{\mathbf{z}(x)}.
}}
\label{MultiSol}
\end{eqnarray}
$\textrm{ for } x_k \leq x \leq x_{k+1}.$ \\
The integration constant $\mathbf{u}$ needs to be determined from the boundary and intersection conditions, demonstrated in subsection \ref{global_solution}.
$ \mathbf{P}(x) $ describes the fundamental system and $ \mathbf{z}(x) $ represents a particular solution of  Eq.~\eqref{balance1}.
The integral in $ \mathbf{z}(x) $ can be solved to 
\begin{eqnarray}
\mathbf{z}(x) =(\mathbf{P}(x) - \mathbf{I}_{8}) \mathbf{M}^{-1}\mathbf{b}
\end{eqnarray}
The solution for an invertible matrix $ \mathbf{M} $, in case of a no-beam simulation, where $ \mathbf{B} $ equals a zero-matrix, is shown later in section~\ref{subsection_Singlegas}. \\
The validity of the solution \eqref{MultiSol} can be easily verified by differentiation.

\subsection{Global solution by implementing boundary and intersection conditions} \label{global_solution}
The local solutions $\mathbf{y}_k(x)$ are now connected with boundary and intersection conditions to form the global solution $ \mathbf{y}(x) $ (similar to expression \eqref{globalSol}) and to determine the integration coefficient $u_k$.  

Some algebraic transformations of the boundary conditions are needed to proceed. We know that:
\begin{eqnarray}
\mathbf{y}(0) = P(0) \cdot  \mathbf{u} +  \mathbf{z}(0) \label{H1} \\
\mathbf{y} (L) = P(L) \cdot  \mathbf{u} +  \mathbf{z}(L)  \label{H2}
\end{eqnarray}
We use $\eqref{H1}, \eqref{H2}$ to transform  $\eqref{IC1}$ to the form
\begin{eqnarray}
\begin{pmatrix}
\mathbf{u}_{k} \\ 1
\end{pmatrix} =  \mathbf{TM}(k-1,k)
\begin{pmatrix}
\mathbf{u}_{k-1} \\ 1
\end{pmatrix},\label{transform1}
\end{eqnarray}
where $\mathbf{TM} \in \mathbb{R}^{9\times 9} $ describes the transformation matrix that maps the unknown $\mathbf{u}$ from segment k-1 to segment k. $\mathbf{TM}$ has the following form for $2 \leq  k \leq N$: 
\\
$ \mathbf{TM}(k-1, k) =  $
\begin{eqnarray*}
	\left( \begin{array}{c|c}
		[ ( \mathbf{H}_k+\mathbf{S}_k)\cdot \mathbf{P}_k(0) ]^{-1} \, \cdot  \,  & \quad -\overline{\mathbf{g}}_k+\mathbf{H}_{k-1} \mathbf{z}_{k-1}(L) \, - \quad \\ 
		\quad \qquad \qquad \mathbf{H}_{k-1}\mathbf{P}_{k-1}(L) & \qquad \qquad (\mathbf{H}_k+\mathbf{S}_k) \cdot \mathbf{z}_k(0) \\
		\\[-1.9ex]
		\hline
		\\[-2.2ex]
		0 \dots \dots 0 & 1
	\end{array}\right)
\end{eqnarray*}
This form can be deduced by the transformation of the intermediate conditions and further elementary algebraic calculations. For more details see Appendix \ref{Appendix}.

We observe with this the following expression that maps the integration coefficient of the first segment to the last segment.
\begin{eqnarray}
\begin{pmatrix}
\mathbf{u}_{N} \\1  \end{pmatrix} = 
\underbrace{\prod \limits_{k=2}^{N} \mathbf{TM}(k, k-1)}_{=: \mathbf{SM}} \cdot 
\begin{pmatrix}
\mathbf{u}_{1} \\1  \label{transform} \end{pmatrix}
\end{eqnarray}
The transformation product defines a new matrix $\mathbf{SM}$.
In the same way we use  $\eqref{H1}, \eqref{H2}$ to modify $\eqref{BC1}$ and $\eqref{EC1}$: \\
Let $ \overline{\mathbf{F}_1}, \overline{\mathbf{F}_N} \in\mathbb{R}^{4\times 9}$ and $\overline{\mathbf{u}_1}, \overline{\mathbf{u}_N} \in\mathbb{R}^{9}$, then we can write:
\begin{eqnarray*}
\overbrace{\left( \begin{array}{c|c}
	\quad \mathbf{F}_1 \mathbf{P}_1(0) \quad &  \quad \mathbf{F}_1 \mathbf{z}_1(0) + \mathbf{g}_1 \qquad
	\end{array}\right)}^{=: \overline{\mathbf{F}_1}}
\cdot \overbrace{\begin{pmatrix}
	\mathbf{u}_{1} \\ 1 
	\end{pmatrix}}^{=: \overline{\mathbf{u}_1}}
&=& \begin{pmatrix} 0\\ 0 \\ 0\\ 0 \end{pmatrix} \\
\underbrace{\left( \begin{array}{c|c}
	\mathbf{F}_N \mathbf{P}_N(L) \quad &  \quad \mathbf{F}_N \mathbf{z}_N(L) - \mathbf{g}_{N+1}
	\end{array}\right)}_{=: \overline{\mathbf{F}_N}}
\cdot \underbrace{\begin{pmatrix}
	\mathbf{u}_{N} \\ 1 
	\end{pmatrix}}_{=: \overline{\mathbf{u}_N}}
&=& \begin{pmatrix} 0\\ 0 \\ 0\\ 0 \end{pmatrix}
\end{eqnarray*}
Note, that the additional vector entry of $\overline{\mathbf{u}_1}$ is required in order to also describe constant algebraic transformations in the boundary conditions.

We rewrite the boundary conditions now to the final system of equations:
\begin{eqnarray}
\underbrace{\left( \begin{array}{c|c}
	\quad \mathbf{SM} \quad & \quad -I_{(9 \times 9)} \quad \\ \hline
	\overline{F_1} & 0_{(4 \times 9)} \\ \hline
	\mathbf{0}_{(4 \times 9)} & \overline{\mathbf{F}_N} \\ \hline
	0 \dots 0 \, 1 & 0 \dots 0 -1
	\end{array}\right)}_{\in\mathbb{R}^{18 \times 18}} \cdot 
\underbrace{\begin{pmatrix}
	\overline{\mathbf{u}_{1}} \\ \overline{\mathbf{u}_{N}} 
	\end{pmatrix}}_{\in\mathbb{R}^{18 \times 1}} =
\underbrace{\begin{pmatrix}
	0 \\ \vdots  \\ \vdots \\ 0 \\0 
	\end{pmatrix}}_{\in\mathbb{R}^{18 \times 1}} \label{final}
\end{eqnarray}
Solving Eq.~$\eqref{final}$ with a Gauss-Jordan elimination algorithm \cite{saad2003iterative}, using the transformation-identity of Eq.~$\eqref{transform1}$ and evaluating $\eqref{MultiSol}$, gives us the solution $\mathbf{y}(x)$. The backward transformation of $\mathbf{y}$ defines the particle density  $\mathbf{n}$ at each axial point of the simulation domain.

\subsection{Special case: Single-gas model} \label{subsection_Singlegas}
The equation system \eqref{ProblemD1} becomes decoupled, if the gas species do not interact with each other. This is the case when the ion-induced desorption matrix $\mathbf{H}_{ion}$ is diagonal, or if $\mathbf{H}_{ion}$ describes the zero-matrix in the case of no beam.

Note, that a diagonal matrix of $\mathbf{H}_{ion}$ can be forced to approximate the solution by the following transformation:
\begin{eqnarray}
\mathbf{\tilde H}_{\textrm{single}}^{\textrm{ion}} = \frac{\sum\limits_{l}\mathbf{H}_{kl}^{\textrm{ion}}\cdot \sigma_l}{\sum\limits_l \sigma_l}
\end{eqnarray}
We solve the equation system individually for each gas species using an exponential approach $n(x) = exp(\lambda x)$ for a $ \lambda \in \mathbb{R}$. This  gives us the following real solutions $n_i(x) $ of Eq.~\eqref{ProblemD1} for the gas specie $i$: $n_i(x) = $
\begin{eqnarray*}
 \left\{
\begin{aligned}
&C_1 \cdot \exp{( \sqrt{-\frac{b}{a}} x)} + C_2 \cdot \exp{(- \sqrt{-\frac{b}{a}} x)} - \frac{c}{b} &  \text{for } b < 0 \\
&C_1 \cdot \cos{(\sqrt{\frac{b}{a}} x)} + C_2 \cdot \sin{(\sqrt{\frac{b}{a}} x)}  - \frac{c}{b} &  \text{for } b > 0\\
&C_1 + C_2  x - \frac{c}{2a}  x^2  &  \text{for } b= 0\\
\end{aligned}
\right. 
\end{eqnarray*}
$ a,b $ and $ c $ are in this case one dimensional coefficients of gas specie $i$ and $ n_i $ is its one-dimensional density function.
The integration constants $ C_1 $ and $ C_2 $ can be easily obtained with the boundary and intersection conditions, following the simplified solution concept from the previous subsection.
\begingroup
\squeezetable
\begin{table}[h]
	\caption{\label{para_math} Model parameters for mathematical description.}
	\begin{ruledtabular}
		\begin{tabular}{llll}
			\textbf{Symbol} & 
			\textbf{Dim} &
			\textbf{Description} \\
			\hline	
			$ \mathbf{y} $ & $\mathbb{R}^{8}$  & vector-valued function describing the RGP density\\
			&& and its derivative\\
			$ \mathbf{I}_{4} $ & $\mathbb{R}^{4 \times 4}$  & Identity matrix  \\
			$ \mathbf{0}_{4 \times 4} $ & $\mathbb{R}^{4 \times 4}$  & Zero matrix  \\
			$\overrightarrow{0}_4 $ & $\mathbb{R}^{4 }$  & Zero vector  \\
			$ 	x $ &  $\mathbb{R}$  & Spatial coordinate along beam line\\
			$ x_k $ &  $\mathbb{R}$ & Intersection point of segment $k-1$ and $k$\\
			$ \mathbf{A}, \mathbf{B} $ & $\mathbb{R}^{4 \times 4}$ &  Coefficients of balance equation  \\
			$  \mathbf{c} $ & $\mathbb{R}^{4 }$   & Coefficients of balance equation  \\
			$ \mathbf{M}$ & $\mathbb{R}^{8 \times 8}$ &  Coefficients of transformed balance equation  \\
			$  \mathbf{b} $ & $\mathbb{R}^{8 }$   & Coefficients of transformed balance equation  \\
			$  \mathbf{P} $ & $\mathbb{R}^{8 \times 8}$   & Fundamental system of transformed balance equation  \\
			$  \mathbf{z} $ & $\mathbb{R}^{8 }$   & Particular solution of transformed balance equation  \\
			$  \mathbf{u} $ & $\mathbb{R}^{8 }$   & Integration constants of balance equation  \\
			$  \overline{\mathbf{u}} $ & $\mathbb{R}^{8 }$   & Integration constants of balance equation with\\
			&& an artificial extra vector entry at the end.  \\
			$ \mathbf{F_1}, \mathbf{F_N} $ & $\mathbb{R}^{4 \times 8}$   & Coefficients of boundary conditions  \\
			$ \overline{\mathbf{F_1}}, \overline{\mathbf{F_N}} $ & $\mathbb{R}^{4 \times 9}$   & Coefficients of homogenized boundary conditions  \\
			$  \mathbf{H}, \mathbf{S} $ & $\mathbb{R}^{8 \times 8}$ &   Coefficients of intersection conditions  \\
			$  \mathbf{H}_{\textrm{ion}},\mathbf{\tilde H}_{\textrm{ion}} $ &$\mathbb{R}^{4 \times 4}$&   Ion- induced desorption matrix wrt multi-\\
			&& and single-gas framework\\
			$ \mathbf{g} $ &  $\mathbb{R}^{4}$  & Local lumped gas source (e.g. leak) \\
			$ \mathbf{s}_k $ & $\mathbb{R}^{4}$  & Pumping speed of lumped pump located at the beginning\\
			&& of segment $k$  \\
			$  \overline{\mathbf{g}} $ & $\mathbb{R}^{8}$ &   Coefficients of intersection conditions  \\
			$  \mathbf{TM} $ & $\mathbb{R}^{9 \times 9}$ &   Transformation matrix: maps coefficients \\
			&&from segment (k-1) to k  \\
			$  \mathbf{SM} $ & $\mathbb{R}^{9 \times 9}$ &   Transformation matrix: maps coefficients from \\
			&&the first segment to the last one.  \\
			$  n(x) $ & $\mathbb{R}$ &   Particle density for one gas specie\\
			&& (Single-gas framework)  \\
			$  a, b, c $ & $\mathbb{R}$ &  Coefficients of balance equation for one gas specie \\
			&&(Single-gas framework), e.g. $a= A_{11}$  \\
			$ C_1, C_2 $ & $\mathbb{R}$ &  Integration constants of balance equation\\
			&& (Single-gas framework)  \\
		\end{tabular}
	\end{ruledtabular}
\end{table}
\endgroup

\section{Validation of the model by benchmark examples}

The model has been thoroughly tested in the framework of benchmark examples and real-case scenarios for the Large Hadron Collider (LHC) at CERN.
We give five representative examples out of the many used as benchmark for this study.

The analytical model presented in this paper is referred to as ``PyVASCO'' in the following section.

\subsection{Crosscheck with Molflow+}
Molflow+ uses a stochastic approach to simulate the RGPs with Test-Particle Monte Carlo methods for one gas specie at a time. Molflow+ traces the trajectory of virtual particles from the gas source to the pumping location and derives from this the RGP density in the vacuum chamber. The advantage of Molflow+ is that it can consider complex geometries. PyVASCO, on the other hand, can consider multiple gas species at a time and beam induced effects. As a side-note, Molflow+ can also consider photon induced desorption by coupling Molflow+ with the closely related program SynRad+ \cite{ady2016monte}.
To meet the assumptions of both models, we choose a simple cylindrical geometry and put the ionization matrix to zero to avoid intermolecular dependencies. We explicitly tested variations of outgassing rates $\mathbf{q}$, sticking coefficients $\alpha$, conductances and diameters. In these benchmark examples both models show a very good agreement. For readability reasons, all figures and tables are inserted at the end of the subsections.

The geometry for the first three examples is visualized in Fig.~\ref{fig:Ex1}. It is one single beam pipe consisting of two materials M$_1$ and M$_2$ defined in Table \ref{material_1}-\ref{material_3}. 
%%%3

\subsubsection{Example - Variation of mass} \label{example1}
The first example represents the influence of the particle's mass to the density distribution. Explicitly, the conductance depends on the mass of the RGP, hence different gas species provide different conductances. Table~\ref{table_mass} lists the molecular masses of $ \ce{H2}, \ce{CH4}, \ce{CO} $ and $ \ce{CO2} $. Fig.~\ref{fig:Plot1} presents then the simulation output for the distribution of the particle density assuming the same outgassing $ \mathbf{q} $ and sticking $ \alpha $ properties for each gas species. The results confirm the well-known fact that the higher  the molecular mass of the specie, the lower is the conductance and the higher is the RGP density.
%%%Table: table_mass 1table + fig:ex1

\subsubsection{Example - Variation of outgassing rate} \label{example2}
This and the following two examples simulate only the density distribution of $\ce{H2}$ particles for the geometry of Fig.~\ref{fig:Ex1}. Example~\ref{example2} focuses on the behaviour of different outgassing values $\widetilde{\mathbf{q}}$ to the simulations. The results are presented in Fig.~\ref{fig:Plot2} and they clearly show a linear relation among the different outgassing coefficients $\widetilde{\mathbf{q}}$. This is an expected result. 
% table and fig

\subsubsection{Example - Variation of sticking coefficient} \label{example3}
Example~\ref{example3} determines the effect of different sticking factors to the density profile.  Fig.~\ref{fig:Plot3} shows that its effect can not be as easily deduced as it was in Example~\ref{example2} for the outgassing rate. The reason is that the amount of particles removed from the system due to the sticking coefficient is depending on the prevailing density, see balance Eq.~\eqref{balanceEQ} from the previous sections.

\subsubsection{Example - Variation of diameter} \label{example4}
\begin{figure}[!h]
	\centering
	\includegraphics[width=0.4\textwidth]{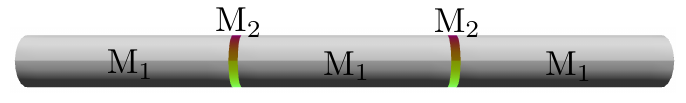}
	\caption{\label{fig:Ex1} Geometry for Example~\ref{example1}-\ref{example3}: One beam pipe with two materials $ \textrm{M}_1 $ and $ \textrm{M}_2 $}	
	\centering
	\includegraphics[width=0.4\textwidth]{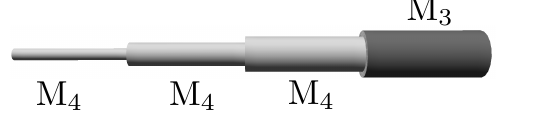}
	\caption{\label{fig:ex2} Geometry for Example~\ref{example4} with materials $ \textrm{M}_3 $ and $ \textrm{M}_4 $}
\end{figure}

\begin{figure}[!h]
	\includegraphics[width=0.4\textwidth]{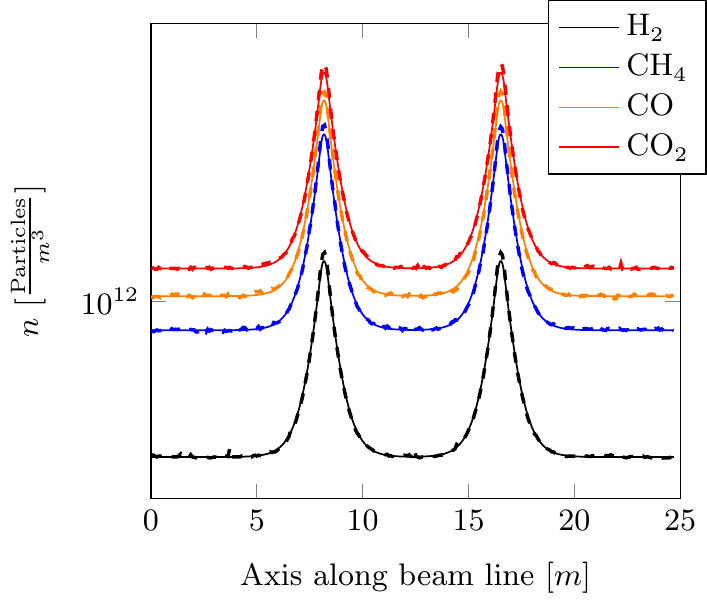}
	\caption{\label{fig:Plot1} Example \ref{example1}: Conductance variation in comparison with PyVASCO~(solid line) and Molflow+~(dashed line) for the geometry of Fig.~\ref{fig:Ex1}.}
\end{figure}

\begin{figure}[!h]
	\includegraphics[width=0.4\textwidth]{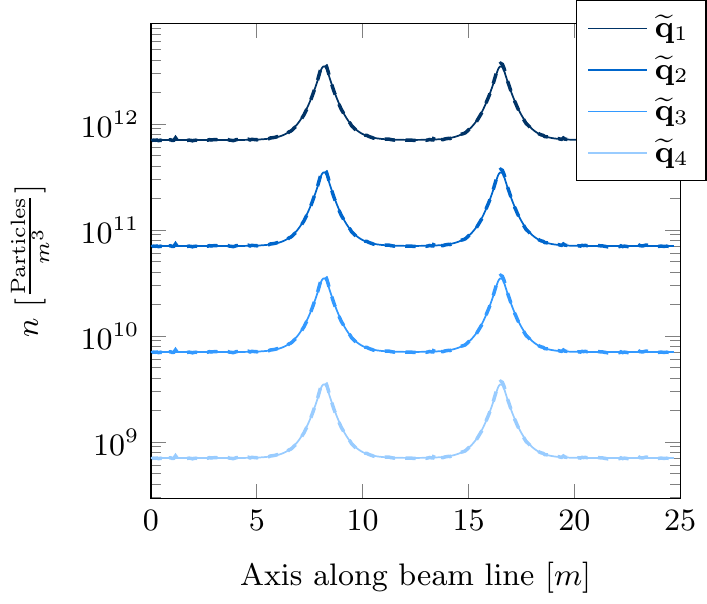}
	\caption{\label{fig:Plot2} Example \ref{example2}: Outgassing variation for $ \ce{H2} $ in comparison with PyVASCO~(solid line) and Molflow+~(dashed line) for the geometry of Fig.~\ref{fig:Ex1}.}
\end{figure}

\begin{figure}[!h]
	\includegraphics[width=0.4\textwidth]{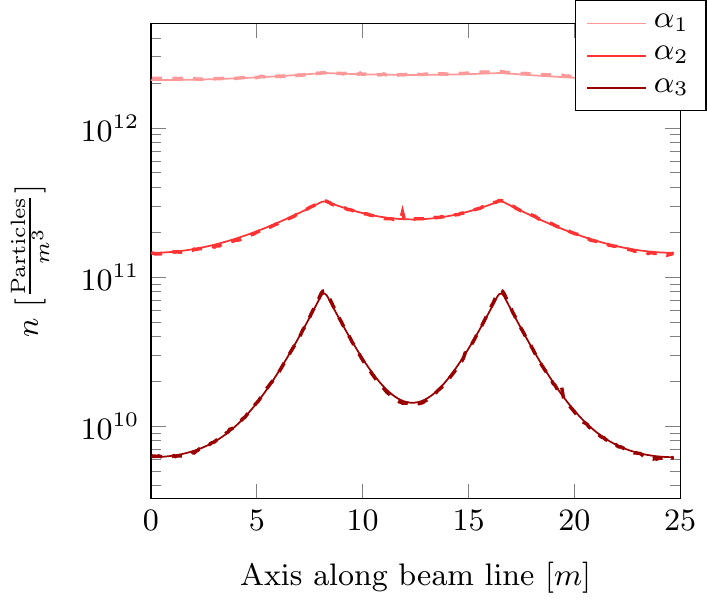}
	\caption{\label{fig:Plot3} Example \ref{example3}: Sticking variation for $ \ce{H2} $ in comparison with PyVASCO~(solid line) and Molflow+(dashed line) for the geometry of Fig.~\ref{fig:Ex1}.}
\end{figure}

\begin{figure}[!h]
	\includegraphics[width=0.4\textwidth]{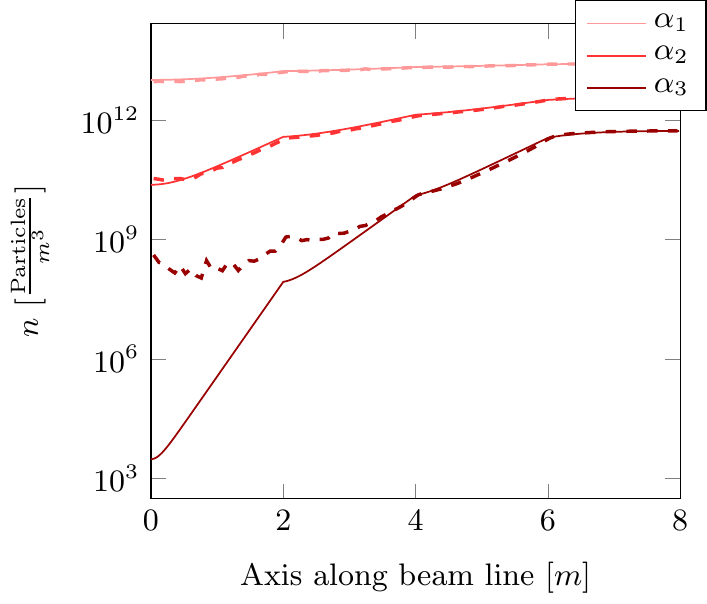}
	\caption{\label{fig:Plot4} Example \ref{example4}: Sticking variation for $ \ce{H2} $ in comparison with PyVASCO~(solid line) and Molflow+ (dashed line) for the geometry of Fig.~\ref{fig:ex2}.}
\end{figure}

Example~\ref{example4} is applied to the geometry of Fig.~\ref{fig:ex2} of material  M$_3$ and M$_4$ listed in Table~\ref{material_4} to test the effect of different geometries. The vacuum chamber is described by four segments with an increasing diameter from 100 - 400mm. 
The results in Fig.~\ref{fig:Plot4} show a good match except for the last case that assumes a very high sticking factor $\alpha \geq 0.1$ and zero outgassing.
We want to note here, that this is an hypothetical test case and there are barely domains in the LHC where this configuration could be found.
The explanation of the underestimation of the density profile is due to the beaming effect. This means, that particles from a gas source  may propagate along the vacuum chamber direction and these particles do not experience the sticking coefficient at all and increase the density in a domain several meters further away. The piecewise solution implemented in our simulation-method cannot capture such an effect \cite{bonucci2007transmission}.

\begin{figure}[!h]	
	\includegraphics[width=0.4\textwidth]{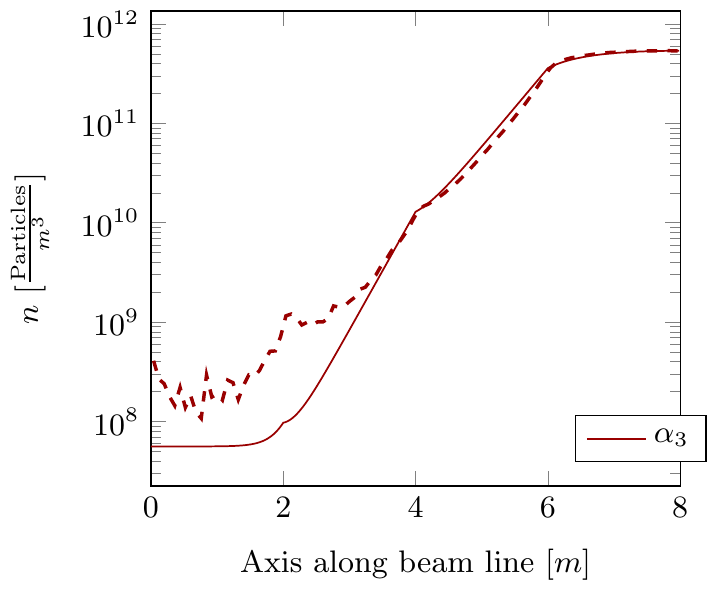}
	\caption{\label{fig:Plot5} Example \ref{example4}: A high sticking factor makes beaming effect visible; comparison PyVASCO~(solid line) and Molflow+ (dashed line) with corrected outgassing $\widetilde{\mathbf{q}}$.}
\end{figure}

A solution to this problem is to set an additional very small outgassing rate $ \widetilde{\mathbf{q}} = 10^{-14} \frac{\textrm{mbar} \cdot \textrm{l}}{\textrm{s} \cdot \textrm{cm}^2} $ in the corresponding domain, here it is set at the first segment that has a diameter of 100mm. The result is plotted in Fig.~\ref{fig:Plot5}. 

\subsubsection{Observations - PyVASCO vs. Molflow+}

All benchmark examples show a good match between the analytic code PyVASCO and the Monte-Carlo code Molflow+. The fact that two models with different approaches give the same result increases the credibility of both simulation codes.

%%%%%%%%%% page with figures
\begingroup
\squeezetable
\begin{table}[!h]
	\caption{Mass of the four main gas species in a UHV system \cite{nistgov}.}
	\begin{ruledtabular}
		\begin{tabular}{l|cccc}	
			\quad &		
			\ce{H2} & 
			\ce{CH4} &
			\ce{CO}&
			\ce{CO2} \\
			\hline
			\textbf{mass}[g/mol] & 2 & 16 & 28 & 44 \\	
		\end{tabular}
	\end{ruledtabular}
	\label{table_mass}
\end{table}
\endgroup 
\begingroup
\squeezetable
\begin{table}[!h]
	\caption{Material $ \textrm{M}_1 $ and $ \textrm{M}_2 $ specifications for Example~\ref{example1}.}
	\begin{ruledtabular}
		\begin{tabular}{l|ll}	
			\quad &		
			$ \alpha $ & 
			$  \widetilde{\mathbf{q}} \times 10^{-12} [\frac{\textrm{mbar}\cdot \textrm{l}}{\textrm{s}\cdot \textrm{cm}^2}] $  \\
			\hline
			$ \textrm{M}_1 $ & $ 8 \cdot 10^{-3} $ & $3.97 $ \\	
			$ \textrm{M}_2  $ & $ 10^{-12} $ & $39.8$  \\		
		\end{tabular}
	\end{ruledtabular}
\label{material_1}
\end{table}
\endgroup 
\begingroup
\squeezetable
\begin{table}[!h]
	\caption{Material $ \textrm{M}_1 $ and $ \textrm{M}_2 $ specifications for Example~\ref{example2} (only $ \ce{H2} $).}
	\begin{ruledtabular}
		\begin{tabular}{l|lllll}	
			\quad &		
			$ \alpha $ & 
			$ \Big[ \widetilde{\mathbf{q}}_1 $ & 	$ \widetilde{\mathbf{q}}_2 $ &	$ \widetilde{\mathbf{q}}_3  $ &	$ \widetilde{\mathbf{q}}_4 \Big] \times 10^{-14} [\frac{\textrm{mbar}\cdot \textrm{l}}{\textrm{s}\cdot \textrm{cm}^2}] $  \\
			\hline
			$ \textrm{M}_1 $ & $ 8 \cdot 10^{-3} $ & $1000 $ &$ 100 $ & $10 $ & $1 $ \\	
			$ \textrm{M}_2  $ & $ 10^{-12} $ & $ 10000 $ & $ 1000 $ &$ 100 $ & $ 10 $  \\		
		\end{tabular}
	\end{ruledtabular}
\label{material_2}
\end{table}
\endgroup 
\begingroup
\squeezetable
\begin{table}[!h]
	\caption{Material $ \textrm{M}_1 $ and $ \textrm{M}_2 $ specifications for Example~\ref{example3} (only $ \ce{H2} $).}
	\begin{ruledtabular}
		\begin{tabular}{l|llll}	
			\quad &		
			$ \alpha_1 $ & $ \alpha_2 $ &	$ \alpha_3  $ &	$ \widetilde{\mathbf{q}}\times 10^{-14}$  \\
			\hline
			$ \textrm{M}_1 $ & $ 10^{-5}  $ & $10^{-4} $ &$ 10^{-3} $  & $8  $ \\	
			$ \textrm{M}_2  $ & $ 10^{-13} $ & $  10^{-12} $ & $   10^{-11}$ & $ 800 $  \\		
		\end{tabular}
	\end{ruledtabular}
\label{material_3}
\end{table}
\endgroup 
\begingroup
\squeezetable
\begin{table}[!h]
	\caption{Material $ \textrm{M}_3 $ and $ \textrm{M}_4 $ specifications for Example~\ref{example4} (only $ \ce{H2} $) .}
	\begin{ruledtabular}
		\begin{tabular}{l|llll}	
			\quad &		
			$ \alpha_1 $ & $ \alpha_2   $& $ \alpha_3   $ &	$ \textrm{Q}$  \\
			\hline
			$ \textrm{M}_3 $ & $  10^{-3} $ & $10^{-2} $ &$ 10^{-1} $  & $10^{-10}  $ \\	
			$ \textrm{M}_4  $ & $ 10^{-3} $ & $ 10^{-2} $ & $ 10^{-1} $ & $ 0 $  \\		
		\end{tabular}
	\end{ruledtabular}
\label{material_4}
\end{table}
\endgroup 
\newpage
\subsection{Sensitivity of the ion-induced desorption}
We study in this subsection the sensitivity of the ion-induced desorption phenomena (see subsection \ref{physical_model}.\ref{flow_into}.\ref{flowinto2}) on the simulation output. A high beam current and a fully occupied matrix $ \mathbf{H}_{\textrm{ion}} $ imply that the different gas-species influence each other and hence mathematically one numerical instability of one gas-specie can map to the other gas species. Hence, an emphasize lies on a stable influence of this phenomena.

\begin{figure}[!t]
	\centering
	\includegraphics[width=0.4\textwidth]{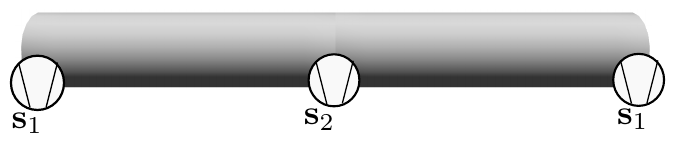}
	\caption{Beam-pipe with three pumps to test the ion-induced desorption sensitivity.}
	\label{fig:Geometry3}
\end{figure}

We study therefore the variation in the output, when we use a multi-gas framework ($\mathbf{H}_{\textrm{multi}} $), a single-gas framework ( $\mathbf{H}_{\textrm{single}} $) and a zero-beam framework ( $\mathbf{H}_{\textrm{zero}} $) with:
\begin{eqnarray*}
	\mathbf{H}_{\textrm{multi}} &=& \begin{pmatrix}
		0.54 & 0.1 &  0.1 &  0.1 \\
		0.1 & 0.54 &  0.1 &  0.1 \\
		0.1 & 0.1 &  0.54 &  0.1 \\
		0.1 & 0.1 &  0.1 &  0.54 \\
	\end{pmatrix}, \\
\mathbf{H}_{\textrm{single}} &=& \textrm{diag}(\mathbf{H}_{\textrm{multi}}), \quad
\mathbf{H}_{\textrm{zero}} = 0 \cdot \mathbf{H}_{\textrm{multi}}
\end{eqnarray*}
Fig.~\ref{fig:Geometry3} shows the 6~m long cylindrical vacuum chamber that we use for this analysis and Table~\ref{pumps} lists the parameter of pumping speed and outgassing rate. The material has no sticking property in this case to illustrate the ion-induced desorption dynamics better.

\begin{figure}[!t]
	\includegraphics[width=0.4\textwidth]{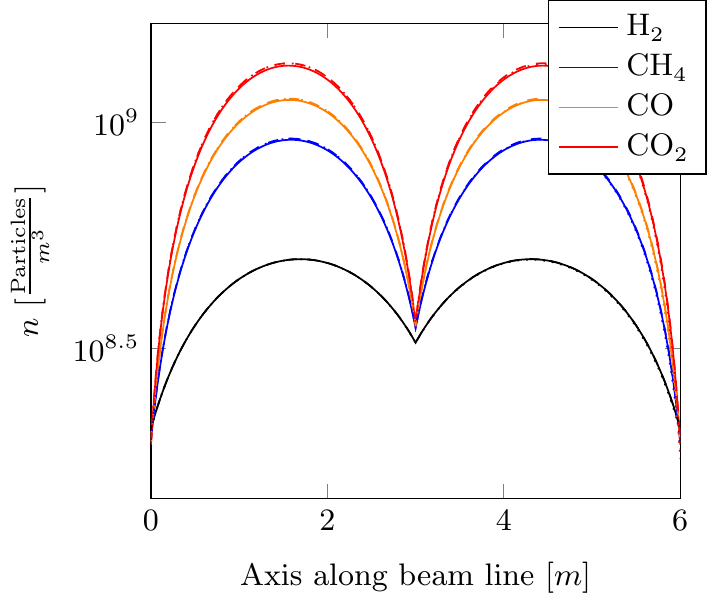}
	\caption{\label{fig:Plot7} Comparison of zero-beam (solid line) and single-gas (dashed line) and multi-gas (dotted line) framework with current $ I= 0.01$A }
\end{figure}

We compare all three frameworks in Fig.~\ref{fig:Plot7}, with the presence of a weak beam ($I=0.01$A). The matrix entries of $\mathbf{H}^{\textrm{ion}}$ are therefore all close to zero and we expect, as shown also in Fig.~\ref{fig:Plot7}, three times the same result.\\

\begin{figure}[!t]
	\includegraphics[width=0.4\textwidth]{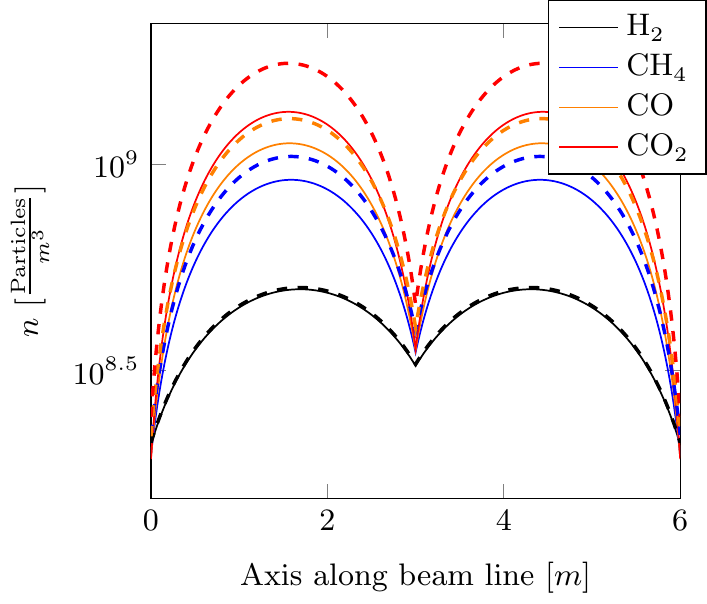}
	\caption{\label{fig:Plot8} Difference between zero beam (solid line) and a high beam current $ I = 10$A (dashed line) for the single-gas framework.}
\end{figure}

In Fig.~\ref{fig:Plot8} we see, how a higher beam current of $I = 10$A influences the output for the single-gas framework; the result is as expected. A higher current implies more collisions of beam particles with RGP, which consequently get ionized.  They impinge on the wall and hence increase the ion-induced desorption. The heavier $\ce{CO2}$ molecules present a higher ionization cross-section than for example $\ce{H2}$. The density increase for a higher beam current is therefore stronger for $\ce{CO2}$ than for $\ce{H2}$.\\

\begin{figure}[!t]
	\includegraphics[width=0.4\textwidth]{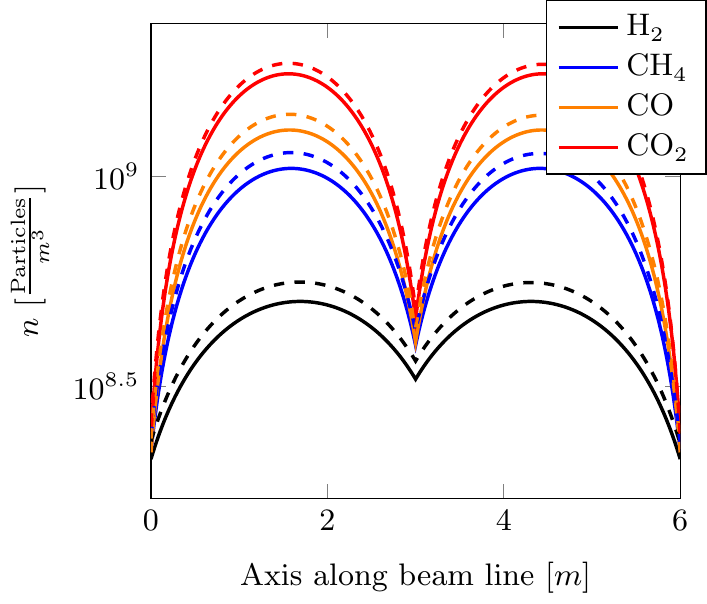}
	\caption{\label{fig:Plot9} Difference between single-gas with $ \mathbf{H}_{\textrm{Single}} $ (solid line) and multi-gas with $ \mathbf{H}_{\textrm{Multi}} $ (dashed line) simulations, both at $ I = 10$A. }
\end{figure}

Fig.~\ref{fig:Plot9} reflects the difference between the multi-gas framework and the single-gas framework at a high beam current ($I = 10$A) and hence especially analyses how the off-diagonal entries influence the result. Firstly, we observe a different values in the density profiles. The additional cross-desorption probability from the off-diagonal entries causes this reasonably small increase. Secondly, the shape of the profile remains the same, as we expect from a stable model, when the input parameters are changed only by a small quantity. Thirdly, $\ce{H2}$ shows the biggest increase in its density. The reason is that the ion-induced desorption is proportional to the prevailing density $\mathbf{n}$. This means that the higher density of $\ce{CO2}$ has a stronger influence on the density of $\ce{H2}$ than inverse.\\

Concluding, the results for testing the sensitivity of the model to the coupled equation term of the ion-induced desorption reflects a stable model and gives no indication to any instabilities. 

%figures:
\begingroup
\squeezetable
\begin{table}[!t]
	\caption{\label{pumps} Parameters for pumps and material properties of Fig.~\ref{fig:Geometry3}.}
	\begin{ruledtabular}
		\begin{tabular}{l|cccc}	
			\quad &		
			\ce{H2} & 
			\ce{CH4} &
			\ce{CO}&
			\ce{CO2} \\
			\hline
			$\mathbf{s}_1$[l/s] &
			1100 &
			1100 &
			1100 &
			1100 \\	
			$\mathbf{s}_2$l/s]  &
			550 & 
			550 & 
			550 & 
			550 \\	
			$\widetilde{\mathbf{q}}  [\frac{\textrm{mbar} \cdot \textrm{l}}{\textrm{s} \cdot \textrm{cm}^2}] \times 10^{-15} $ & 
			1 &
			1 & 
			1 & 
			1\\
			\textbf{$ \alpha $}  & 
			0 & 
			0 & 
			0 & 
			0\\
			\textbf{$ \sigma \times 10^{-23} $} & 
			4.45 & 
			31.8 & 
			27.5 & 
			42.9 \\
		\end{tabular}
	\end{ruledtabular}
\end{table}
\endgroup

\section{Validation of the model in comparison to LHC gauge readings}
The last step to validate the model lies in the comparison of the simulation results with measured data from gauges in the LHC. The LHC containes eight long straight sections (LSS) consisting of two parts of about 265~meters with a point of interest (detector) in the middle. The CMS experiment is installed in one LSS  and we compare the pressure gauge readings in this area for a beam of 6.5~TeV  with our simulations. 

The simulations contains a combination of many different density regimes driven by the 14 varying input of  geometries, materials and beam induced effects.  We divided the domain into 464 segments on which we assumed constant values for the parameters. The segments are connected with the intersection conditions \eqref{IC} described previously.

\begin{figure}[t]
	\centering
	\includegraphics[width=0.5\textwidth]{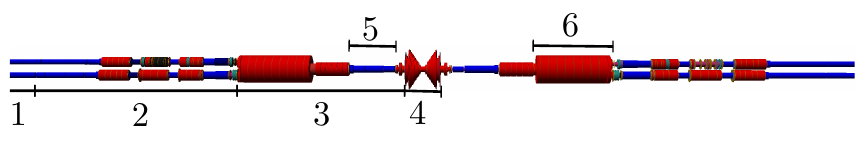}
	\caption{Sketch of the CMS experimental area:
		1:~Arc, 2:~Dispersion surpressor, 3:~Long Straight Section, 4:~CMS experimental area, 5:~Inner triplet, 6:~Recombination chamber }
	\label{fig:CMS_sketch}
\end{figure}

The collision point is at the centre of the simulation domain. This area is dedicated to provide the lowest possible gas density.  The components of the vacuum chamber on both sides of the collision point are roughly symmetric: The inner triplets, with high magnetic gradients to focus the beam for the collisions; the normal-conducting separation magnets, that splits the vacuum chambers in two separate parts for each beam; the so called dispersion suppressor that is connected then to the arcs (see Fig.~\ref{fig:CMS_sketch}).

\subsection{Geometry and material assumption}
The LHC database  stores all the parameter specifications of the vacuum chamber \cite{layoutdatabase, cdd, bruning2004lhc}. For our calculations, we have extracted from the database the diameter,  length and material specification, that affect the conductance, thermal outgassing and sticking probability.

\subsubsection{Arc}
The bending arcs consist of a repetitive structure of three times 15m long chambers for dipole magnets and of shorter 6m long chambers for focusing magnets and beam instrumentation measurements. The vacuum chambers or so-called cold bores are joined by a short stainless steel bellow.
The diameter of the cold bores are small about 40-60mm and additionally, a racetrack-shaped beam screen is implemented inside the chamber \cite{cruikshank1997mechanical, }. To avoid multiple reflections of the photons on the wall a so-called sawtooth surface is indented on the internal side of the beam screen, where the primary synchrotron radiation photons hit. The sawtooth profile is characterized by a reduced photon reflectivity which allows localisation of the molecular desorption and of the photoelectrons (see e-cloud effect in subsection~\ref{flowinto3}). In the strong magnetic regions (as in the arcs) the cold-bores of the superconducting magnets are cooled by liquid helium at 1.9-4.5 K. They act as distributed cryogenic pumps, via the pumping slots on the beamscreen.

\subsubsection{Transition area}
The vacuum chambers of the straight sections are generally kept at room temperature. All transitions from cryogenic to room temperature  chambers happen within half a meter containing usually a valve, a vacuum module with beam instrumentation equipment and flanges at their extremities. These chambers are connected with bellows that compensate thermal expansions due to the large temperature gradients during the commissioning of the system.

\subsubsection{Straight Section}
The vacuum chambers in the straight sections are at room temperature and are principally made out of copper with an additional NEG coating, in order to reduce photon-induced desorption and the generation of secondary electrons. In between NEG coated parts, there can be short higher outgassing parts found, e.g. vacuum modules with beam instrumentation or beam collimation equipment. Collimators are often located before sensitive instruments, close to the detectors and at the end of magnet assemblies to intercept stray particles.
The most critical area, the vacuum chamber in the centre of CMS, is made out of beryllium. This material has a higher radiation length than copper to provide a higher transparency to, and lower absorption of, the exotic particles resulting from the beams' collisions. 
Generally, the diameter in the straight section varies from 80 to 230mm. It reaches its maximum aperture in the recombination chamber, where two beam lines combine into one common chamber.  

\subsubsection{Material properties}
The realistic  outgassing rates and desorption coefficients for the materials are estimated on the basis of laboratory results measured by the BVO section of the vacuum group\cite{Giuseppe}  or by reference values from literature \cite{chiggiato2014vacuum, baglin2002synchrotron, baglincoupled}, \cite{bruning2004lhc, nistgov}. However, material treatment for ultra-high vacuum, like vacuum firing, bake out, activation and beam-conditioning are special treatments, and therefore parameters may vary in time and from standard values in the literature.

\subsubsection{Beam induced parameters}
\begin{figure}[b]
	\includegraphics[width=0.4\textwidth]{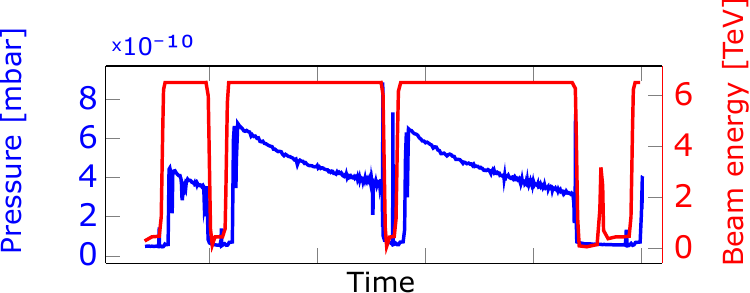}
     	\caption{\label{fig:timeserieschartimagetime25092809} Time dynamics of gauge VGI.220.1R5.X (blue curve) in the common beam chamber close to CMS from 25.09.2016 to 28.09.2016, derived from \cite{timber} and compared to the time-dynamics of the beam energy (red curve).}
\end{figure}

During the operation, beam induced effects are the predominant factor that influence the RGP density and hence increase it by orders of magnitude. Fig.~\ref{fig:timeserieschartimagetime25092809} shows the correlation between the dynamics of the beam energy and the readings of one specific pressure gauge.

Photon and electron induced desorption take place especially in the magnetic areas of the LHC, which present about $ 90 \% $ of the total accelerator.

\begin{figure}[t]
	\centering
	\includegraphics[width=0.4\textwidth]{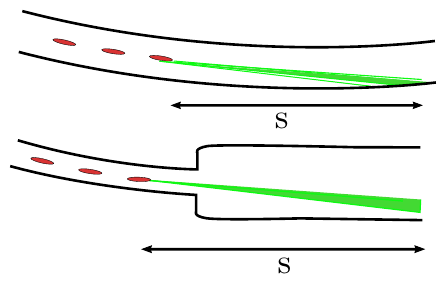}
	\caption{Distance s between the generation of a photon and its impingement on the wall in the relativistic case.}
	\label{fig:SR_Travel}
\end{figure}

Formula~$\eqref{photonflux} $ implies a photon flux generated by the dipoles in the arcs of
\begin{eqnarray}
\dot{\Gamma}_{ph} = \frac{1.5414 \cdot 10^{21}}{2 \rho \pi} = 8.77 \cdot 10^{16}  \textrm{Photons/(m s)} 
\end{eqnarray}
where $\rho = 2795.84$~m, $ I = 0.5 $~A and $ E = 6.5 $~TeV.
Additionally, we consider that synchrotron radiation emitted by the beam travels tangentially to the orbit, deducing a spread between the location of the  generation of the photons and their impingement on the chamber wall. This distance can be up to 20 meters and it depends on the geometry of the beam chamber (see Fig.~\ref{fig:SR_Travel}). 

This implies two consequences for the input parameters: Firstly, there may be photons impinging on the following chambers downstream of the radiation source points even if there is no magnetic field present. Thus,  we must take into account a longitudinal shift of the impinging photon flux at the appropriate spot in our simulations. Secondly, a change in the geometry may cause a high spike of impinging photons, when a chamber is smaller than the previous one.
Beam collimators for example  provide a small variable chamber aperture to remove the beam particles in the halo and consequently are subjected to a high photon bombardment \cite{valentino2013beam}. Additionally, synchrotron radiation is also produced by an off-axis beam in quadrupoles and orbit correctors. However the magnetic strength of these magnets is much less compared to the bending magnets in the arcs. The critical energy $E_c$ for dipoles in the LHC can be  estimated with 40eV, whereas the strong focusing quadrupoles before the experiments provide a maximum of 9eV. \\
Summarizing, the distribution of the photon flux in the arcs is characterized by a more or less continuous distribution, whereas in the straight sections it consists of distinct peaks.

The electron cloud is strongly depending on the chamber diameter, the material, bunch spacing and the beam energy and intensity. 

Moreover, two beams are present in the common chambers of the LHC, that leads to a significant beam-induced density increase especially in the quadrupole triplets close to the detector.

In addition, the study of the evolution of the heat dissipation $ Q $ on the vacuum chambers may also give a hint for values of the beam induced parameters. These parameters are as well logged in \cite{timber}. It holds that:
\begin{eqnarray}
Q_{\textrm{heat}} = Q_{\textrm{SR}} + Q_{\textrm{IC}} + Q_{\textrm{EC}},
\end{eqnarray}
where SR refers to synchrotron radiation, IC to image current and EC to electron cloud.

\subsection{LHC gauges}
There are a total of 98 gauges installed in the area of CMS that monitor the vacuum dynamics. The final goal now is to compare the simulation output with the readings of the installed gauges. Most of the LHC gauges are inverted-magnetron penning gauges (IKR 070, Pfeiffer), which mainly measure down to $10^{-11}$mbar. However for beam lifetime and radiation background reasons, the pressure in the LHC is in some parts lower than this value, and therefore they serve mainly as an alarm system that indicates potential vacuum degradation. This explains the rather high gauge reading in the NEG-coated recombination chamber in Fig.~\ref{fig:cmspartc0differencewhitetotal}, located about 100~m left and right from the CMS interaction point. Bayard-Alpert ionization gauges (SVT 305) are also employed. These measure down to values of the order of $10^{-12}$mbar \cite{pigny2015measurements}. 

\begin{figure}[b]
	\centering
	\includegraphics[width=0.9\linewidth]{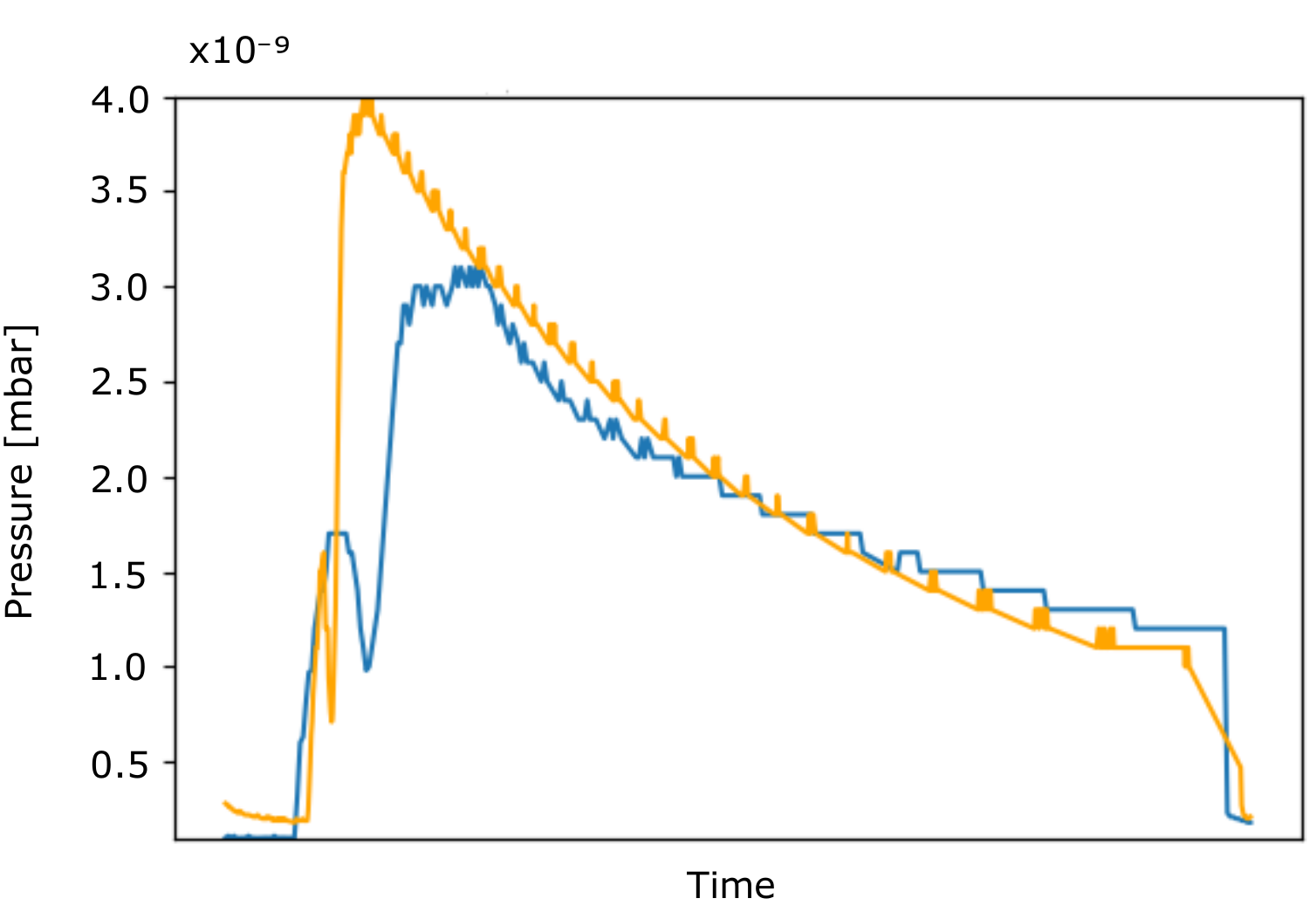}
	\caption{Time evolution of penning gauge VGPB.242.7L5.B for Fill 5211 (orange line) and Fill 5338 (blue line).}
	\label{fig:13073}
\end{figure}

For our discussions here, we derived the gauge's readings of four very similar runs from mid-August to the end of October in 2016 indicated with  a fill number of 5211, 5338, 5416 and 5451 from the LHC logging database \cite{timber}. In Fig.~\ref{fig:13073} the time evolution of the gauge VGPB.242.7L5.B is plotted for the Fill 5211 and 5338. This graphic should visualize that fills with similar parameters provide similar results. The slight difference is due to a higher monitored emittance for beam of Fill 5211.
The monitored pressure value of the gauges is nitrogen equivalent and therefore the gauges' sensitivity to different gas species must be taken into account in our calculations. Unfortunately, the harsh radiation environment of the LHC tunnel does not allow the installation of residual-gas analysers and their delicate electronics.

\subsection{Results and discussion}
\begin{figure*}[bt]
	\includegraphics[width=0.8\textwidth]{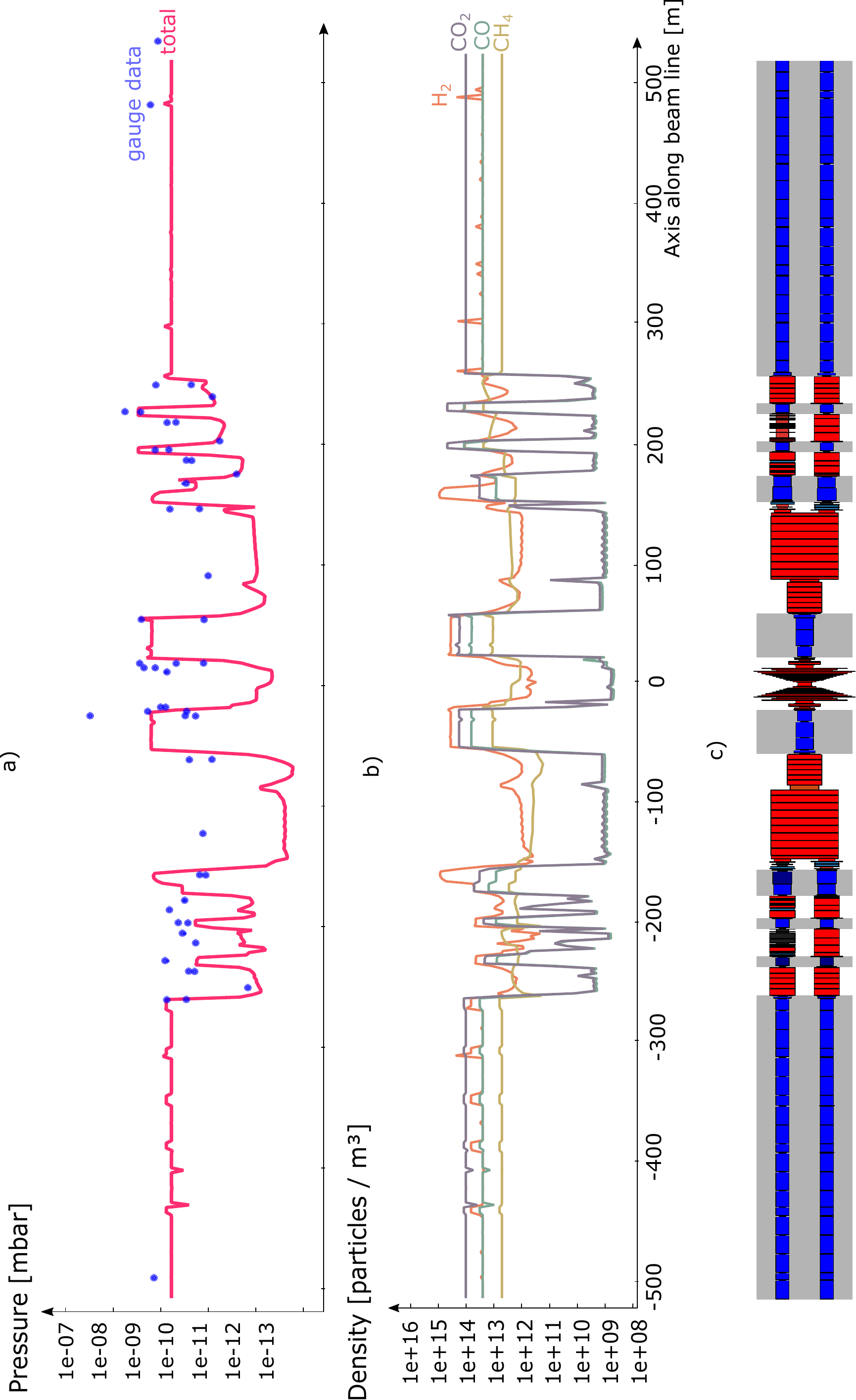} 	%[width= 0.8\textheight, 	height= 1.3\textwidth,]
	\caption{Vacuum simulation of the Long Straight Section with the CMS detector located in the middle: a)~total pressure plot, b)~density plot of the four gas species and c)~a geometry sketch.}
	\label{fig:cmspartc0differencewhitetotal}
\end{figure*}

The RGP density in the experimental area around CMS provides many different characterizing aspects, that are visualized as our main result in Fig.~\ref{fig:cmspartc0differencewhitetotal}.\\

\begin{figure}[t]
	\centering
	\includegraphics[width=0.9\linewidth]{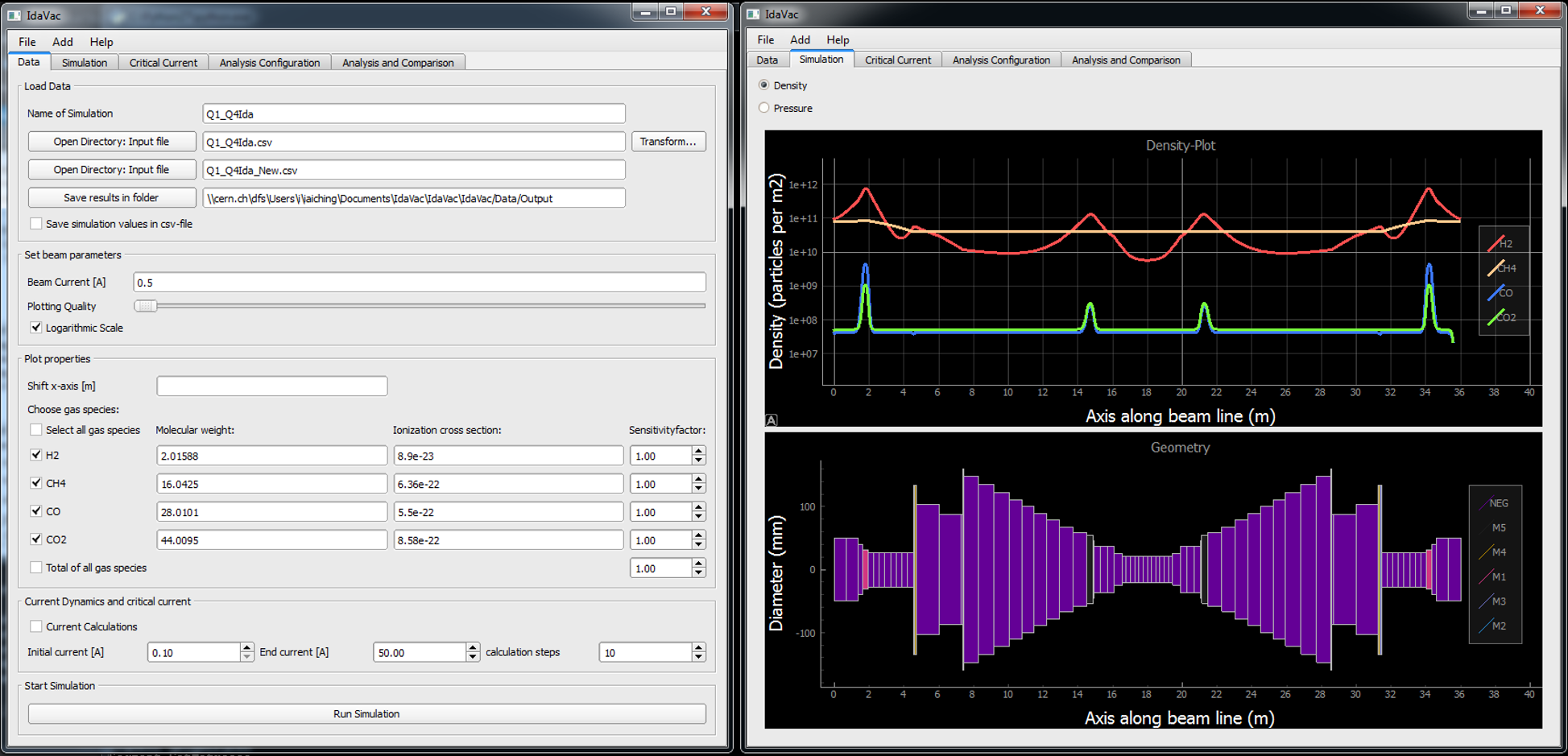}
	\caption{Screenshot of the simulation program ``PyVASCO''. }
	\label{fig:picture1}
\end{figure}

We implemented the model in a Python environment and embedded it in a graphical user interface based on the library of PyQt \cite{meier2015python, newman2013computational, lutz2013learning, hill2016learning} (see Fig.~\ref{fig:picture1}). Results are calculated within less than a minute.

Fig.~\ref{fig:cmspartc0differencewhitetotal} shows that the general goal is fulfilled, that the maximum density in the LHC should not exceed $ 10^{15} \ce{H2}$-equivalent gas particles per $ \textrm{m}^3 $ in the presence of the circulating beams. The RGP density in the experimental area is even by orders of magnitude lower to minimize the background noise to the experiments. Hence, a beam lifetime in the order of 100 hours supports an efficient operation of the high energy experiments with respect to vacuum requirements. 

The four different graphs in Fig.~\ref{fig:cmspartc0differencewhitetotal} represent the prevailing densities of $ \ce{H2}, \ce{CH4}, \ce{CO} $ and $ \ce{CO2} $. Their different shapes symbolize their different behaviours. Hydrogen constitutes the major part of the gas load. Hydrogen's low mass and low binding energy compared to the other gas species results in a higher probability of beam induced desorption and a higher thermal outgassing of the vacuum chamber walls. 

The shape of the plot of $ \ce{H2}, \ce{CO} $ and $ \ce{CO2} $  show a similar structure. $\ce{CO} $ and $ \ce{CO2} $ are of the same order of magnitude. The biggest difference of each gas specie's value can range of three orders of magnitude.

The different materials are represented by different colours in the geometry sketch of CMS in Fig.~\ref{fig:cmspartc0differencewhitetotal}. As an outline, cryogenic areas are marked in dark blue with a grey background, room temperature areas are marked in red.

The biggest density gradient can be observed in the transition areas from room temperature to cryogenic area. The following relation holds:
\begin{eqnarray}
\frac{n_1}{n_2} = \sqrt{\frac{T_2}{T_1}}
\end{eqnarray}
This tells us that the RGP density in areas with lower temperatures is higher than in areas with higher temperatures, assuming the same material specifications. Additionally, the high density has to be related with beam induced effects that occur mainly in the arcs. The density profile also visualizes the importance of the beam direction. The density is slightly higher, when the beam travels from the arcs to the Straight Section, because of the fact that emitted photons can travel several meters until they imping on the wall (see Fig.~\ref{fig:SR_Travel}).

The density plot for hydrogen shows some peaks in the arcs. They appear exactly at the interconnects between two vacuum chambers. In this part the liquid helium cooling pipe on the beam screen is usually absent for a few centimetres, which leads to a slight temperature rise and hence the hydrogen equilibrium density goes up.

Otherwise, cryo-  and NEG pumping results in a flat line, at the equilibrium of surface pumping and degassing. Lumped pumps are only located in room-temperature areas.  

Thin-film NEG coating deposited along all room-temperature chambers, capture getterable gas species such as $ \ce{H2}, \ce{CO} $ and $ \ce{CO2} $. This configuration provides very efficient distributed surface pumping. Methane is, due to its closed-symmetric atomic structure, not reacting with the surface and is not pumped \cite{chiggiato2006ti}. This is the main reason, why the density profile of methane is clearly different from the other gas species.
Methane is only pumped by lumped ion pumps. Its pressure profile therefore resembles a parabola from one pump to the next. This can be seen very well  between the quadrupoles Q4 and Q7 in Fig.~\ref{fig:cmspartc0differencewhitetotal} about 250~m right of the CMS-interaction point.  Its should be added that there are indications that $\ce{CH4}$ is also pumped by beam-ionisation, and therefore if this effect is not taken into account the density curves for this gas are to be intended as worst case scenario \cite{mathewson1996beam}.
The low RGP density in the collision area is due to the outgassing characteristics of beryllium and the extremely low photon bombardment in this area.

\section{Conclusion}
This article introduces a mathematical model and a computer code to calculate and efficiently forecast the residual gas particle density in a particle accelerator. This quantity should be kept as low as possible to support an efficient operation of the machine. Several effects influence it and make this requirement challenging. Among them are beam-induced effects as well as thermal outgassing, diffusion inside the chamber and interactions among the different residual gases. The idea was to combine all of them in one mathematical model which gives as output the density distribution of the four dominating gas species $ \ce{H2}, \ce{CH4}, \ce{CO} $ and $ \ce{CO2} $ inside a beam pipe.  A mass-balance equation system of second order serves this purpose. Based on mathematical theorems a solution was found and the fundamental steps to this goal have been shown.

The validation of the model was established by a cross check with the Test-Particle Monte Carlo code Molflow+, a sensitivity analysis of the ion-induced desorption term including a comparison between the single-gas and the multi-gas framework, and finally by a comparison to  gauge readings in the LHC. The latter cross-check was presented for the Long Straight Section close to the CMS experiment. All these simulations show reasonably meaningful results and consequently suggest realistic replication of the vacuum environment. The knowledge on how, where and why these values influence the vacuum quality provides consequently a great aid in the design and analysis of vacuum systems. In addition, the results are computed rather fast in less than 30 seconds, even for large simulation domains.

This model provides the potential to undergo detailed parameter variation studies, hence to understand the main influencing effects at different locations and therefore to detect critical configurations in advance that could lead to vacuum degradation. 

Nowadays, CERN's new challenge is to develop concepts for post-LHC circular particle colliders (FCC) \cite{benedikt2014future} and the next step is to use this simulation model to present a variety of possible designs and to choose among them, in agreement with further specifications, the best possible solution.

\begin{acknowledgments}
The discussions with Giuseppe Bregliozzi, Josef Sestak and Vincent Baglin were most helpful to find appropriate input parameters. Many thanks as well to Jan Sopousek, who helped the authors with the implementation of the model in a Python environment and to Marton Ady, who supported the authors with Molflow+ simulations.
Thanks also to Adriana Rossi for discussions about the previous model VASCO.
All authors work in the Vacuum Surfaces and Coatings group at CERN. This project and its achievements are part of the global future circular collider study hosted by CERN. Ida Aichinger is a doctoral student at the Johannes Kepler University Linz, Austria, supported by the Austrian Doctoral Student Programme of CERN. 
\end{acknowledgments}

\appendix
\onecolumngrid
\section{Algebraic transformation of intermediate boundary condition} \label{Appendix}
The  value of the unknown $ y $ is posed once at the end of segment k-1 and once at the beginning of segment k. For the posed solution function $y$ by the Theorem of Picard Lindel\"{o}f, we always need to know the value of $y$ at the beginning of each segment. \\
We start with the initial conditions
\begin{eqnarray*}
	H_{k-1} y_{k-1}(x_k) - (H_k+S_k) y_k (x_k) = G_k
\end{eqnarray*}
and use the identity $ \eqref{H1} $ and $ \eqref{H2} $, we receive:
\begin{eqnarray*}
H_{k-1} \Big( P_{k-1}(L)\cdot y_{0(k-1)} +q_{k-1}(L)\Big) - (H_k+ S_k ) \cdot \Big( P_k(0) \cdot y_{0k} + q_k (0) \Big) = G_k \\
(H_k+S_k) \cdot \Big( P_k(0) \cdot y_{0k} + q_k(0) \Big) = -G_k + H_{k-1} \underbrace{\Big( P_{k-1} (L) \cdot y_{0(k-1)} + q_{k-1} (L) \Big)}_{ =: ( \ast )}  \\
P_k(0) \cdot y_{0k} + q_k(0) = - (H_k+S_k)^{-1}G_k + \underbrace{(H_k + S_k)^{-1} H_{k-1} (\ast)}_{=: (\ast\ast)} \\
y_{0k} = -P_k^{-1} (0) (H_k+S_k)^{-1}G_k + P_k^{-1} (0) \cdot (\ast\ast)- P_k^{-1} q_k (0)
\end{eqnarray*}
\begin{eqnarray*}
y_{0k} = \stackrel{{=:M\in\mathbb{R}^{8\times 8}}}{\boxed{ P_k^{-1}(0) \cdot (H_k+S_k)^{-1}H_{k-1}P_{k-1}(L)}} y_{0(k-1)} + \\
\stackrel{{=:v\in\mathbb{R}^{8}}}{\boxed{ - P_k^{-1}(0)(H_k+S_k)^{-1}G_k + P_k^{-1}(0)\cdot (H_K+S_k)^{-1}H_{k-1}q_{k-1}(L) - P_k^{-1} q_k (0)}} \\
\quad \\
\Rightarrow 
\begin{pmatrix}
y_{0k} \\ 1  \end{pmatrix} = 
\left( \begin{array}{c|c}
M & v \\
\hline
0 \dots 0 & 1 \end{array}\right) \cdot\
\begin{pmatrix}
y_{0(k-1)} \\ 1  \end{pmatrix}
\end{eqnarray*}
Some more algebraic transformations give finally:
\begin{eqnarray*}
	y_{0k} = \underbrace{\Big[(H_k +S_k) \cdot P_k(0) \Big]^{-1}}_{M_2} \cdot  
	\Big[ \underbrace{H_{k-1} P_{k-1} (L)}_{M_1} \cdot y_{0(k-1)} - \underbrace{G_k + H_{k-1}q_{k-1}(L) - (H_k+S_k)\cdot q_k(0)}_{\tilde{V}} \Big]
\end{eqnarray*}
\begin{eqnarray*}
	\begin{pmatrix}
		y_{0k} \\ 1 \end{pmatrix} =
	\underbrace{
		\left( \begin{array}{c|c}
			M_2^{-1}M_1 & \tilde{V} \\
			\hline
			0\dots 0 & 1 \end{array}\right)}_{ = TM(k, k-1)} \cdot 
	\begin{pmatrix}
		y_{0(k-1)} \\ 1  \end{pmatrix}
\end{eqnarray*}
\twocolumngrid
\bibliography{IdaAichingerPRAB_references}

\end{document}